\documentclass{acm_proc_article-sp}
\usepackage{graphicx}
\usepackage{times}
\usepackage{balance}
\usepackage{amssymb}
\usepackage{array}
\usepackage{url}
\usepackage{graphicx}
\usepackage{booktabs}
\usepackage{epstopdf}
\usepackage{float}
\usepackage{enumerate}
\usepackage{multirow}
\usepackage{algorithm}
\usepackage{amsmath}
\usepackage{algpseudocode}
\usepackage{color}
\usepackage{caption}
\usepackage{subfigure}
\usepackage{etoolbox}
\usepackage{times}
\usepackage{textcomp}
\makeatletter
\def\@copyrightspace{\relax}
\makeatother

\begin{document}
\bibliographystyle{abbrv}
\title{When Human Cognitive Modeling Meets PINs:\\ User-Independent Inter-Keystroke Timing Attacks}

\author{
	Ximing~Liu\textsuperscript{1}, Yingjiu~Li\textsuperscript{1}, Robert~H.~Deng\textsuperscript{1}, Bing~Chang\textsuperscript{1},
	Shujun~Li\textsuperscript{2}\\ \affaddr{\textsuperscript{1}School of Information Systems, Singapore Management University, Singapore}\\
	\affaddr{\textsuperscript{2}Kent Interdisciplinary Research Centre in Cyber Security (KirCCS) and School of Computing, University of Kent, UK}\\
	\email{\{xmliu.2015, yjli, robertdeng, bingchang\}@smu.edu.sg; S.J.Li@kent.ac.uk}
}
\maketitle

\begin{abstract}
This paper proposes the first user-independent inter-keystroke timing attacks on PINs. Our attack method is based on an inter-keystroke timing dictionary built from a human cognitive model whose parameters can be determined by a \textit{small} amount of training data on any users. Our attacks can thus be potentially launched in a large scale in real-world settings. We investigate inter-keystroke timing attacks in different online attack settings and evaluate their performance on PINs at different strength levels. Our experimental results show that the proposed attack performs significantly better than random guessing attacks. We further demonstrate that our attacks pose a serious threat to real-world applications and propose various ways to mitigate the threat.
\end{abstract}


\section{Introduction}
Inter-keystroke timing attacks, which make use of the leaked keystroke timing information to infer a user's PIN, pose a serious threat to real-world applications, especially for online financial services whose authentication systems are based on PINs. Such attacks have triggered increasing interests in recent years due to the development of many practical approaches to obtaining users' keystroke timing information via different side channels, including CPU cache~\cite{gruss2015cache, oren2015spy, pessl2016drama, gruss2016flush+, lipp2016armageddon}, shared event loops~\cite{vila2017loophole}, I/O interrupts~\cite{diao2016no, lipp2017practical, zhang2009peeping}, and SSH~\cite{song2001timing}. Some approaches do not even require attackers to be physically close to victims or install malware on victims' devices, which significantly lower the barrier for launching inter-keystroke timing attacks in real-world scenarios.

Most of the existing inter-keystroke timing attacks on PINs or passwords are user-dependent. Since the seminal work published by Dawn Song et al.\ in 2001~\cite{song2001timing}, the Hidden Markov Model (HMM) has been exploited as a major technique to launching the inter-keystroke timing attacks~\cite{zhang2009peeping, KuneKim2010TimingPIN}. However, HMM is user-specific in a sense that it relies on the distribution of inter-keystroke times of a specific user typing each possible key pair (which represents a hidden state in HMM) so as to infer the user's PIN from the user's inter-keystroke timing information about a PIN entry. In other words, HMM requires that a sufficiently large amount of time intervals for each possible key pair that can be part of any PIN be typed by a specific user for model training so as to make the attacks to that specific user's PIN entry accurate and useful. It is usually difficult for an attacker to collect such large amount of inter-keystroke data about a victim before launching an effective attack. Even if it is possible, such attacks are not scalable. If an attacker intends to compromise a new victim, he/she needs to collect the new victim's inter-keystroke timing data about all possible key pairs and retrain his/her HMM for the new victim. In addition, the success rate of such attacks is too low to be practical in online attack settings since the number of guesses that is allowed to launch an online attack is usually restricted to small numbers (e.g., 3, 10, 100) in common practice.

In this paper, we propose a user-independent approach to exploit inter-keystroke timing information for PIN inference, which makes inter-keystroke timing attacks much more scalable and practical. The model in our attacks is not user specific, which can be trained from a small amount of training data (e.g., a few key pairs instead of all possible key pairs) about \textit{any} users (e.g., attackers themselves or people recruited by attackers) instead of the target victim. In addition, our approach can be applied to attack \textit{any} new victim without retraining the model. The success rate of our attacks is significantly higher than random guessing attacks, which poses a serious threat when applied to users in a large scale, even in online attack settings.

Our proposed approach leverages a human cognitive model to capture the common characteristics across \textit{all} skilled users typing PINs. The human cognitive model is derived from several PIN typing behavioral phenomena which we summarize from the cognitive psychology literature. These PIN typing behavioral phenomena are universal to \textit{all} skilled users. The parameters of our cognitive model can be estimated by a few key pairs from any user such as the attacker himself. Once the cognitive model is built, it can be used to attack any user inputting any PIN on a particular keypad whose geometric measurement is known.

At a high level, our attacks proceed as follows. First, an attacker builds a timing dictionary including all possible PINs and their corresponding timing sequences. The timing sequence of each PIN is derived from our cognitive model. Second, the attacker obtains the timing sequence of a victim's PIN entry via various side-channels (e.g., CPU cache, shared event loops, I/O interrupts, and SSH). Third, the attacker measures the cosine similarity between the observed inter-keystroke timing sequence and each entry in the timing dictionary and ranks all candidate PINs in the dictionary by their similarity values. Lastly, with a ranked list of candidate PINs, the attacker may launch online attacks using the PINs successively from the ranked list until he/she succeeds or the target account is locked (or the attacker aborts before the account is locked).

Besides the cognitive model that captures the common characteristics across all users typing PINs, another contributing factor to the user independence of our approach is the way an attacker measures the differences between a victim's PIN entry and each time sequence in the timing dictionary. We adopt cosine similarity since it is invariant to scaling of input vector. It can thus mitigate the negative impact of different typing speeds by different users.

We discover that the effectiveness of our attacks is different for different types of PINs. To examine the effectiveness of our attacks to different types of PINs, we study the inner structure of the whole PIN space and partition the PIN space into different strength levels. In particular, the 6-digit PIN space is partitioned into 5 PIN strength levels according to the directional density of the inter-keystroke timing sequences in the timing dictionary, where \textit{level 1} is the weakest and \textit{level 5} is the strongest. Our attacks achieve much better performance on the PINs at the first four levels compared to the strongest level (i.e., \textit{level 5}). For example, the attacks with 100 guesses on the PINs at \textit{levels 1, 2, 3, 4} are 869, 221, 250 and 42 times more effective than on the PINs at \textit{level 5}, respectively. The results suggest that users should choose their PINs at the strongest strength level for better security in the presence of inter-keystroke timing attacks.

We seek various ways to improve the success rate of our attacks. One question is whether we can achieve better performance using target victims' data for model training, which is commonly used in the existing inter-keystroke timing attacks. In this case, we train the cognitive model using the victim's own inter-keystroke timing data and launch our attacks to this victim's PIN entry. However, the results show that this way improves the success rate by about 4\% only. Another question is whether an attacker can improve his/her success rate if he/she observes the victim's PIN entry for multiple times. In this case, the attacker can attack based on the average of the inter-keystroke timing sequences for multiple PIN entries from the same victim. It achieves around 2\% performance improvement which is not significant either. Our study in these two cases shows that our approach is user-independent and it does not improve much using user-dependent data.

We further examine the scenario in which an attacker happens to know the values of certain digits of the target PIN before launching inter-keystroke timing attacks. It is reasonable to assume that an attacker may attain such knowledge about PIN digits due to the existence of many side-channel attacks (e.g.,~\cite{de2016identification, yue2014blind, shukla2014beware, liu2015good, wang2016Friend, sun2016visible}) and shoulder surfing attacks~\cite{tari2006comparison} to the PIN entry. Unsurprisingly, the success rate of our attacks is significantly improved due to the shrink of timing dictionary in our attacks. For example, when the attacker knows 2 digits, the success rate of attacking the PINs in \textit{level 1} within 3 attempts is improved to 34.9\% so that one out of every three target users can be successfully compromised. In this case, our attacks are practical in online attack settings when applied to a single user or a small number of users.

In general, the success rate of the proposed attacks may not be sufficiently high to pose imminent danger to an individual user's PIN. Our attacks are still practical because they are user-independent and can be applied to attack any number of users in a large scale. To show this, we study two cases in practical settings, where PINs are used as the only credential to protect users' accounts and where attackers can collect many users' inter-keystroke timing data for PIN entries using malicious JavaScripts.

The first case is an online banking system which has three million users. If ten percent of its users' inter-keystroke timing data about PIN entries were collected, our online attacks can be applied to all these users' accounts with 50 tries per account, which do not lead to any account being locked in practice. Consequently, more than 12,000 users' accounts would be compromised on average and these users' account balances and other sensitive information such as usernames and addresses are leaked.

The second case is a mobile payment platform which has more than 520 million users, where each user's ID of his/her account is a mobile phone number. It is not difficult for attackers to obtain many users' mobile phone numbers (e.g., from public web pages or online markets). The login attempts of each user's account in this platform is limited to 3. On the average, an attacker needs to launch online attacks to 83 users' accounts in order to compromise one account. In other words, if our attacks were applied to 1/1000 of users' accounts, then 6,000 users' accounts would be compromised on the average. Our attacks would cause serious damages since attackers can transfer money from victims' accounts to other accounts. Both cases show that our attacks pose a serious threat to real-world applications when applied in large scale.

To mitigate our attacks, we provide several solutions, including choosing longer PINs, choosing PINs at the strongest strength level, proposing a new keypad layout design, and implementing leakage resilient password systems (LRPSes). For the first countermeasure, the security strength of most existing PIN systems is determined by the success probability of random guessing attacks~\cite{Yan2012LRPS}. However, the security strength of PIN systems would be lowered significantly in the presence of our inter-keystroke timing attacks. We suggest users to choose 10-digit PINs to maintain the same security strength under our attacks as that of the 6-digit PINs under random guessing attacks. This solution does not require any change to the hardware of current PIN systems, though it requires users to remember longer PINs.

To relax the requirement on PIN length, our second suggestion is that users should choose PINs at the strongest strength level (i.e., \textit{level 5} for 6-digits PINs\footnote{\textit{Level 5} includes 900,000 PINs which account for 90\% of the total 6-digit PINs. It is thus relatively easy for a user to obtain a PIN at \textit{level 5} if he/she simply chooses his/her PIN randomly.}). Our study on 6-digit PINs shows that the success rate of attacking PINs at \textit{level 5} is around 10 times higher than random guessing attacks. Therefore, to achieve the similar security strength of 6-digit PIN under random guessing attacks, we suggest users choose 7-digit PINs at the strongest strength level.

For the third countermeasure, if changes can be made to the keypad layout, we propose a novel keypad design secure against our proposed attack, which is also easy to use. Our new design nullifies all inter-keystroke timing attacks, which means the success rate of our attacks would be similar to that of random guessing attacks. Therefore, users can still use 6-digit PINs as before. For the last countermeasure, LRPSes have been well studied in the past two decades. A recent work~\cite{Yan2012LRPS} shows that in order to achieve the same security strength of current 6-digit PIN systems, LRPSes require hundreds of seconds to complete an authentication session~\cite{hopper2001secure, asghar2010new}, which sacrifices their usability.

\section{Preliminaries}
In this section, we provide the basics about how to collect inter-keystroke timing information from users, how to model users' typing behavior and what our adversary model is.

\subsection{Keystroke Timing Collection}
\label{datacollection}
To launch any inter-keystroke timing attacks, an attacker needs to collect inter-keystroke timing information about users' inputs. Several practical approaches have been proposed in recent years on how to collect inter-keystroke timing information through various leakage channels, including CPU cache~\cite{gruss2015cache, oren2015spy, pessl2016drama, gruss2016flush+, lipp2016armageddon}, shared event loops~\cite{vila2017loophole}, I/O interrupts~\cite{diao2016no, lipp2017practical, zhang2009peeping}, and SSH~\cite{song2001timing}.

The first leakage channel through which attackers can collect inter-keystroke timing information is CPU cache~\cite{gruss2015cache, oren2015spy, pessl2016drama, gruss2016flush+, lipp2016armageddon}. Through CPU cache, an attacker can observe the effects of a user's keystroke operations and deduce the timestamp of each keystroke the user performs on a keyboard. One of these approaches~\cite{oren2015spy} can be performed from browser sandboxes through remote websites using JavaScripts instead of installing malware on users' devices. Besides users' keystroke operations originated from hardware keyboard, other interactive operations, such as tap operations and swipe operations which are usually triggered on a touch screen, can also be monitored by an attacker~\cite{lipp2016armageddon}. Therefore, inter-keystroke timing attacks (including ours) can be applied to both devices with hardware keyboard and devices with soft keyboard. This keystroke timing collection approach requires malware installed on victim's device to access CPU cache but it does not need any permission.

The second leakage channel through which attackers can collect inter-keystroke timing information is shared event loops~\cite{vila2017loophole}. Through shared event loops in Google Chrome, an attacker can scan an event-delay trace using JavaScript and deduce the timestamp of each keystroke the user performs on a keyboard. This keystroke timing collection approach requires an attacker to trick victims to open a malicious website which has the permission of running JavaScript.

The third leakage channel through which attackers can collect inter-keystroke timing information is I/O interrupts~\cite{diao2016no, lipp2017practical, zhang2009peeping}. An attacker may continuously acquire timestamps using JavaScript in a measuring process and monitor differences between subsequent timestamps~\cite{lipp2017practical}. Significant time differences will occur whenever the measuring process is interrupted by I/O operations (i.e., keystroke operations). The exact timestamp where the user presses a key is clearly visible and can be distinguished from other events. This keystroke timing collection approach also requires an attacker to trick victims to open a malicious website which has the permission of running JavaScript.

The last leakage channel through which attackers can collect inter-keystroke timing information is SSH~\cite{song2001timing}. Since every individual keystroke typed by a user is sent to a remote machine in a separate IP packet immediately after the key is pressed, precise inter-keystroke timings of the user's keystrokes can be learned from the arrival times of the packets. This keystroke timing collection approach requires an attacker to monitor the network and collect the arrival time of SSH packets which does not require any malware to be installed on victim's device or any permission from the victim.

The sampling rates of inter-keystroke timing information collected by different approaches are different (e.g., 40,000Hz for shared event loops and 100Hz for SSH). In our experiments, we use JavaScript to record the key code of each keystroke event and the corresponding timestamp to get the ground truth. We observe that the timings of key-press events are distributed in clusters with a gap of 15 or 16 milliseconds; thus, the sampling rate in our experiments is no higher than $1000/15\approx 66.7$Hz. Although our sampling rate is relatively low, our attacks still achieve satisfactory performance as shown in our experimental results (Section~\ref{performance}). The performance of our attacks may be improved further at higher sampling rates.

To determine the start and the end of victim's PIN entry, the attacker can monitor all the packets sent by the victim by a network sniffing tool on network packets such as Wireshark and records the timestamps of all packets whose destination IP is the targeted sensitive website (e.g., online banking website or Alipay)~\cite{li2016csi}. Since most of the important websites and applications are secured via HTTPS, it does not protect the meta data of the traffic such as destination server's IP address, which can be used to recognize the start of a time window for searching the victim's PIN entry using various approaches which have been mentioned earlier in this section. If the victim is entering PIN on an Android application, Cheng et al.~\cite{cheng2017study} proposed a no-root approach to detect login activities as they share a common pattern that a login activity usually consists of two EditText fields for inputting a username, and a password and the second EditText field sets the attribute inputType to password-related by developers. In addition, malware installed in the victims' phone may make use of accessibility feature to monitor the timing of any event that is activated by the victim by the id of the view~\cite{Accessibility}. Since most developers use EditText fields with an id of `password' or `PIN' in the layout view, it is easy for the attacker to know the start time of a victim's PIN entry event.

\subsection{Human Cognitive Models}
\subsubsection{History}
Human cognitive models have been studied in the field of psychology for decades. They describe one or more specific human cognitive processes (e.g., memory, perception, attention, reasoning, and problem-solving) for the purpose of better understanding, predicting and simulating human behavior~\cite{anderson2013architecture}.

Typing PINs on a numeric keypad is one of the most important human-computer interactions in our daily life and it involves complicated interactions of concurrent perceptual, cognitive, and motoric processes~\cite{wu2007queueing}. To model typing behaviors and explore its underlying mechanisms, cognitive psychologists apply the knowledge of psychology, human-computer interaction and neuroscience. Card et al.~\cite{card1980keystroke} propose a keystroke-level model (KLM) to predict the time of a user accomplishing a given task without errors using a given interactive computer system. For typing task, KLM gives a rough estimate of the average inter-keystroke time, which is calculated by dividing the total time taken in a typing test by the total number of non-error keystrokes. Rumelhart and Norman~\cite{rumelhart1982simulating} build a model of typing and provide detailed predictions about the movement of fingers and the relative response time for letters in different contexts. Furthermore, John~\cite{john1996typist} proposes a typing performance theory which is built within the framework of the Model Human Processor (MHP)~\cite{card1986model} and can offer a more precise estimation. These models of cognitive processing have provided a wealth of information regarding how humans interact with keyboards.

Cognitive psychologists and HCI researchers have also developed several software tools for estimating human performance in terms of time needed by an average skilled user to finish a specific task, such as Cogulator~\cite{Cogulator}, CogTool~\cite{teo2008cogtool}, SANLab-CM~\cite{patton2010sanlab}. Such tools are normally used for modeling and simulating complicated processes involving both computer and human users, but this paper focuses on determining the parameters of a specific model of the typing behavior, so we do not use such tools in our work. In the following sections, we build a new keystroke model combining models mentioned above with empirical analysis.

\subsubsection{Typing Behavior Phenomena}
\label{phenomena}
Typing is a complex procedure involving cognitive activities as well as body movements, but we can still capture the common characteristics across all skilled users' typing behaviors. The typing procedure usually involves two parts: (1) \textit{cognition} of the task and (2) \textit{motor} of the task. During the cognition process, the user conducts a memory retrieval process. Specifically in our scenario, the user recalls his/her PIN from the long-term memory, stores it into the working memory and mentally prepares for executing physical actions. During the motoric process, the user moves his/her hand and fingers to the right key, presses the key, releases the key and prepares for the next keystroke. The total time between two keystrokes is the sum of the time for these two parts.

PIN entry behavior is one of the most common typing behaviors in our daily life. In order to explore PIN entry behavior, we generalize four typing behavior phenomena. They are based on the literature (e.g.,~\cite{salthouse1986perceptual}) which discusses several common phenomena about typing behaviors across all skilled users.

\textit{Phenomenon 1.} \textit{The rate of typing is dependent on how familiar the user is with the typed string.} According to a statistics report, 46 percent of the U.S. students use credit cards on a regular basis for everyday purchases~\cite{Statista}. And the average iPhone user tends to unlock his/her device 80 times in a day~\cite{iPhone} while Android users tend to unlock their smartphones an average 110 times a day~\cite{Android}, so there is no doubt that people are proficient in typing their PINs.

\textit{Phenomenon 2.} \textit{The variability of inter-keystroke time decreases with an increase in users' skill level.} The distributions of inter-keystroke time for the same keystroke in the same context but across multiple repetitions are similar~\cite{salthouse1984effects}. This phenomenon indicates that the typing pattern will stabilize after several practices.

\textit{Phenomenon 3.} \textit{Inter-keystroke time of typing decreases following the power law of practice.} Typing speed of a user will be significantly improved as the number of inputs increases. According to the learning curve of the single user in the study of Gentner~\cite{gentner1983acquisition}, the improvement of inter-keystroke time follows exponential growth. If a skilled user can input PINs smoothly enough, the time of cognition process may be negligible. One reason is that muscle memory has been built after frequently typing and it may take little time for the cognitive processor to make decisions and schedule actions with the motor processor.

\textit{Phenomenon 4.} \textit{The inter-keystroke time is dependent on the specific context, especially for the topography of the keyboard.} The specific context here refers to the character before and after the target character. This topographical effect has been reported by Gentner~\cite{gentner1982evidence, gentner1983acquisition}, Rumelhart and Norman~\cite{rumelhart1982simulating}, and Shaffer~\cite{shaffer1973latency}. Intuitively, the latency between two keystrokes has a positive correlation to their distance on the keypad.

Based on Phenomena 1 and 2, the action of entering a PIN can be regarded as conducted by a skilled user whose typing pattern is stable and predictable. Based on Phenomena 2 and 3, we arrange a practice session before data collection in our experiments in order to collect skilled users' data and simulate people entering PINs in real life. For Phenomenon 4, we estimate the topographical effect by a function of the finger's moving distance and the size of target key using Fitts's law~\cite{fitts1954information}.
\begin{figure} [!t]
	\centering
	\includegraphics[width=0.8\linewidth]{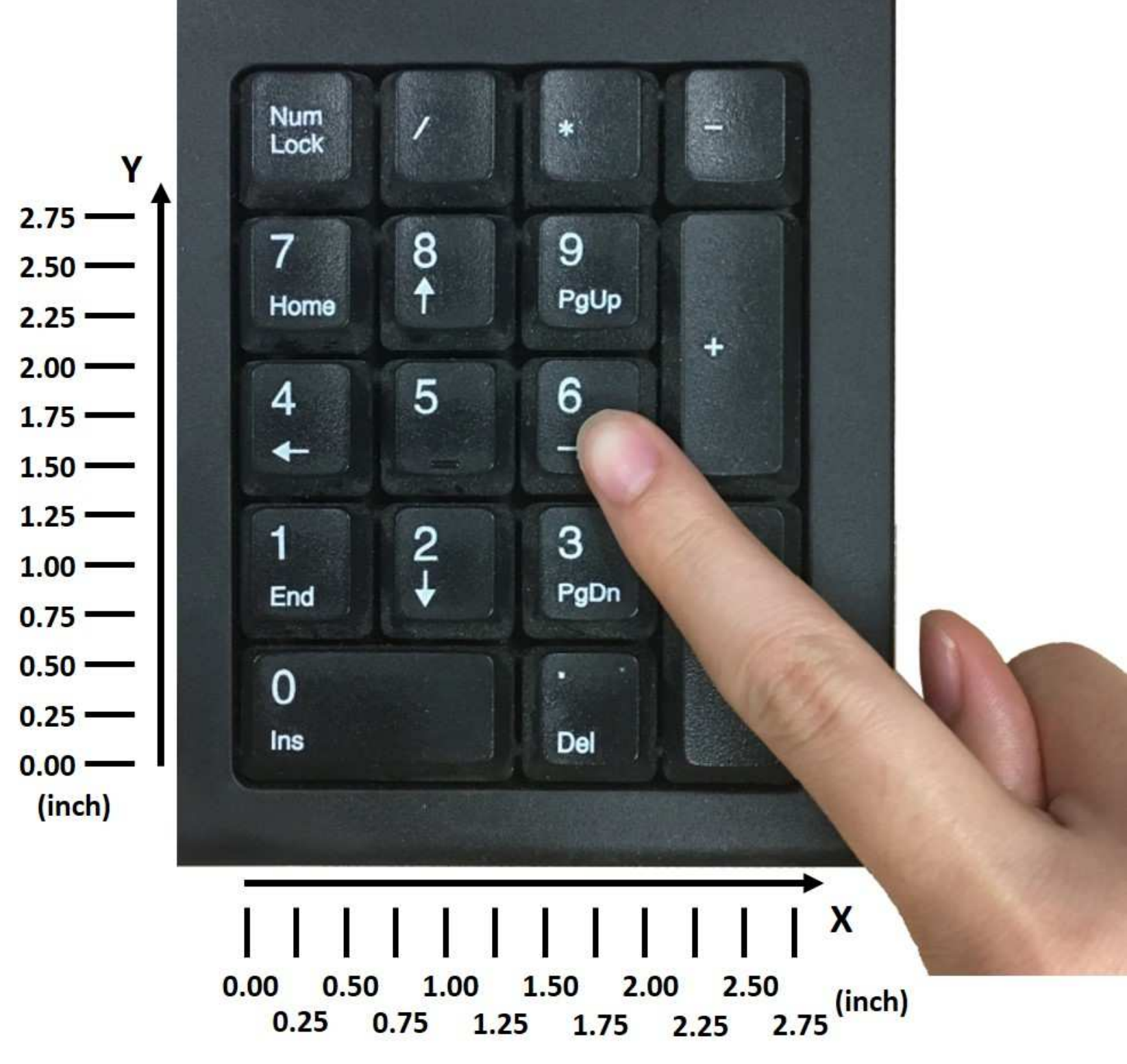}
	\caption{The layout of the numeric pad used in our experiments.}
	\label{fig:NumericPad}
\end{figure}

\subsubsection{Inter-Keystroke Timing Modeling}
\label{timedic}
We incorporate the above typing behavior phenomena to construct a linear model for predicting the inter-keystroke timings of any key pair.

For the topographical effect, our model uses Fitts's law~\cite{fitts1954information} to make finer predictions. Fitts's law is a descriptive model of human movement which can predict the time required to move to a target area. It is used to model the act of physically touching an object with a finger or virtually pointing to an object. Striking the numeric keypad with one finger can be seen as this kind of action. It is a function of the ratio between the distance to the target ($D$) and target width ($W$):
\begin{equation}
T = a + b * I = a + b * \log_{2}(\frac{D}{W} + 1),
\end{equation}
where $D$ is the distance from the start point to the center of the target, $W$ is the effective width of the target in the direction of the motion\footnote{According to our observations in the experiments, the effective press area of each key is close to a circle centered on the center of the key and with a radius equal to the shorter side of the key (which is 0.5 inches). Therefore, we use 0.5 inches as the effective width for all keys including 0 and $<$Enter$>$ keys.}, $I = \log_2(D/W+1)$ is called the \textit{index of difficulty}, $a$ and $b$ are parameters varying from context to context.

We use the geometric center of each key to obtain the distance of each key pair. As for the repeated pressed key like `99', we set $I=0$ so that $T_{motor}=a$. We estimate the values of $a$ and $b$ using inter-keystroke timing data of real human users. With these inter-keystroke timing data and the geometric measurement of victim's keyboard, an attacker can build his/her own inter-keystroke timing model.

We also examine other cognitive operations which may affect the inter-keystroke times, including word-segment effect and word-end effect. We extend the cognitive model and conduct a significance testing on all coefficients to validate the model. However, the results of significance testing show that the impact of word-end effect and word-segment effects are statistically insignificant (see Appendix~A). We thus decide to consider the topographical effect only in our cognitive model.

\subsection{Adversary Model}
\begin{figure*}
	\centering
	\includegraphics[width=\linewidth]{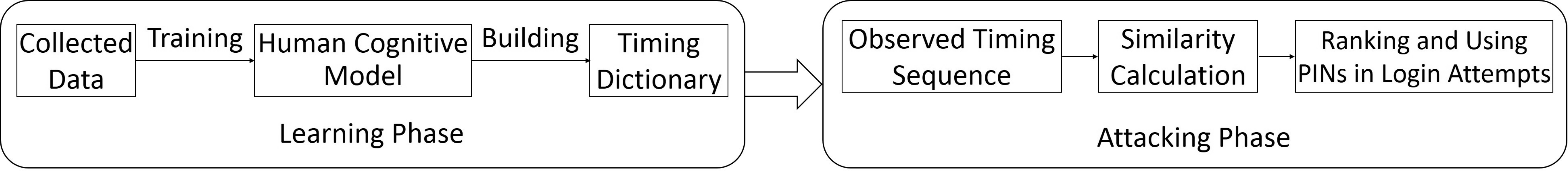}
	\caption{Overview of our inter-keystroke timing attacks.}
	\label{fig:overview}
\end{figure*}
\subsubsection{Basic Premises}
It is usually difficult for a malware to directly record keystrokes due to the use of keylogger detection technologies~\cite{ortolani2010bait, sukhram2017keystroke, tian2017online, ortolani2012noisykey, aslam2004anti}. The barrier for launching inter-keystroke timing attacks in real-world is much lower than directly recording keystrokes. Recent works (e.g.,~\cite{gruss2015cache, oren2015spy, pessl2016drama, gruss2016flush+, lipp2016armageddon,vila2017loophole,diao2016no, lipp2017practical, zhang2009peeping,song2001timing}) have introduced many practical approaches to attaining user's keystroke timing information. While these works focused on how to capture keystroke timing information, our work focuses on how to make use of keystroke timing information to recover PINs. Therefore, our adversary model assumes that an attacker has already obtained the inter-keystroke timing information about a target user (victim) typing his/her PIN on a numeric keypad.

The inter-keystroke timing information about a PIN can be observed just once or a number of times via several leakage channels such as CPU cache, shared event loops, I/O interrupts, SSH as introduced in Section \ref{datacollection}. We notice that directly recording keystrokes requires certain permissions which are usually difficult to be gained (e.g., most software keyloggers require Windows hooks); in comparison, it is relatively easier for an attacker to obtain keystroke timing information. In particular, the shared event loop approach~\cite{vila2017loophole} and the I/O interrupts approach~\cite{lipp2017practical} require that victims' browsers support JavaScript, which is common for popular browsers in the default setting. In addition, the CPU cache approach~\cite{lipp2016armageddon} and the SSH approach~\cite{song2001timing} require no permission to obtain keystroke timing information.

It is also assumed that an attacker knows the layout of the keyboard (including the size of each key and the distance between each key pair) which the target victim uses in advance. This is a reasonable assumption since in most cases, the victim inputs his/her PINs on the number pads of ATMs, POS terminals, or standard keyboards. The layouts of these keypads are standardized or can be easily obtained in the public place. Figure~\ref{fig:NumericPad} shows the layout of a DELL SK-8115 numeric keypad which is used in our experiments.

For the victim's PIN typing behavior, it is assumed that one finger is used to enter the whole PIN followed by an $<$Enter$>$ key press to signal the end of a PIN entry process. It is also a reasonable assumption since according to our observation and the survey\footnote{Please refer to the Section~\ref{limitation} for the detailed statistical results.} we conducted during the experiments, a majority of users (62\%) prefer using a single finger for PIN entry.

\subsubsection{Online Attacks}
In online attack settings, an attacker consecutively tries a number of candidate PINs to attack a PIN-protected account until the correct PIN is found or the account is locked (or the attacker aborts before the account is locked). The online attack that has been studied in most previous research on PIN systems~\cite{martinovic2014authentication} is random guessing attack in which an attacker inputs random PINs. In this paper, we consider four other online attack settings by assuming that an attacker has different knowledge about a victim's typing behavior or the target PIN:

\begin{enumerate}[(i)]
\item \textit{General attacks}: An attacker collects a small amount of inter-keystroke timing data from the attacker himself or people he/she recruits for model training and obtains a single inter-keystroke timing sequence of a PIN entry made by a victim for PIN inference.
\item \textit{Targeted attacks}: An attacker collects a small amount of inter-keystroke timing data about a victim typing known numerical sequences for model training and obtains a single inter-keystroke timing sequence of a PIN entry made by the victim for PIN inference.
\item \textit{Multi-entry attacks}: An attacker collects a small amount of inter-keystroke timing data from the attacker himself or people he/she recruits for model training and captures several inter-keystroke timing sequences about a victim entering the same PIN. In this case, the attacker may combine all inter-keystroke timing sequences and obtain an averaged timing sequence for PIN inference.
\item \textit{Known digits attacks}: An attacker knows certain digits of a target PIN before launching our \textit{general attacks}. Such knowledge may be attained from various side-channel attacks~\cite{de2016identification, yue2014blind, shukla2014beware, liu2015good, wang2016Friend, sun2016visible} or shoulder surfing attacks~\cite{tari2006comparison}.
\end{enumerate}

\noindent\textbf{Limited number of login attempts.} Most PIN systems enforce suspicious login detection and lockout~\cite{freeman2016you}, and thus the number of PINs an attacker may try in an online attack is limited. A successful online attack is defined as an attacker hitting the correct PIN within the number of allowed attempts. The number of login attempts is normally restricted to 3 for PINs with payment cards. When a payment card is used on a POS terminal or with a card reader, entering a PIN wrongly for 3 times may get the card locked. This limit is usually larger for mobile devices. For example, an Android device gets locked temporarily for 30 seconds after every 5 failed attempts, while an iOS device is restored to factory settings after 10 failed logins. Other cases limit online attackers to no more than 100 consecutive failed attempts on a single account according to the digital authentication guidelines~\cite{Digitalauthentication} and electronic authentication guidelines~\cite{Electronicauthentication}. In our experiments, we demonstrate the success rates of our attacks with various limits on the number of consecutive login attempts.

\subsubsection{Offline Attacks}
In offline attack settings, it is assumed that an offline validation of guessed PINs can be performed. This is a less realistic scenario since users' PINs are usually stored in tamper-resistant hardware security modules on the server side as a common practice. Therefore, we focus on the online attacks in this paper.

\section{Attack Methodology}
In this section, we describe the steps of our inter-keystroke timing attacks in detail. Figure~\ref{fig:overview} shows an overview of our inter-keystroke timing attacks, including a learning phase and an attacking phase. In the learning phase, an attacker trains a cognitive model based on certain collected data and builds a timing dictionary. In the attacking phase, the attacker (i) observes one or more entries from a victim, (ii) calculates the similarity between the timing sequence of the observed PIN entry and the calculated timing sequence of each entry in the timing dictionary, (iii) ranks all candidate PINs according to the similarity values, and (iv) attempts to login to the victim's account using the PINs in the ranked list starting from the top in an online attack.
\begin{table*} [!htbp]
	\centering
	\small
	\caption{A segment in the inter-keystroke timing dictionary used in our experiments.}
	\begin{tabular}{|p{2.2cm}<{\centering}|p{2.2cm}<{\centering}|p{2.2cm}<{\centering}|p{2.2cm}<{\centering}|p{2.2cm}<{\centering}|p{2.2cm}<{\centering}|p{2.2cm}<{\centering}|}
		\hline
		PINs   & $K_1$-$K_2$ & $K_2$-$K_3$ & $K_3$-$K_4$ & $K_4$-$K_5$ & $K_5$-$K_6$ & $K_6$-$<Enter>$ \\ \hline
		504316 & 232.9502    & 232.9502    & 237.2201    & 231.3787    & 237.2201    & 226.0874        \\ \hline
		504317 & 232.9502    & 232.9502    & 237.2201    & 231.3787    & 231.3787    & 268.5020        \\ \hline
		504318 & 232.9502    & 232.9502    & 237.2201    & 231.3787    & 237.2201    & 256.9941        \\ \hline
		504319 & 232.9502    & 232.9502    & 237.2201    & 231.3787    & 250.0087    & 247.2787        \\ \hline
		504320 & 232.9502    & 232.9502    & 237.2201    & 199.0121    & 203.7241    & 244.2814        \\ \hline
		504321 & 232.9502    & 232.9502    & 237.2201    & 199.0121    & 199.0121    & 254.0817        \\ \hline
		504322 & 232.9502    & 232.9502    & 237.2201    & 199.0121    & 135.9120    & 232.9502        \\ \hline
		504323 & 232.9502    & 232.9502    & 237.2201    & 199.0121    & 199.0121    & 203.7241        \\ \hline
		504324 & 232.9502    & 232.9502    & 237.2201    & 199.0121    & 214.2976    & 259.6575        \\ \hline
		504325 & 232.9502    & 232.9502    & 237.2201    & 199.0121    & 199.0121    & 243.2131        \\ \hline
	\end{tabular}
	\label{timingdictionary}
\end{table*}

\subsection{Learning Phase}
\noindent\textbf{Data Collection.}
In the learning phase, an attacker needs to collect the inter-keystroke timing sequences for a small number of key pairs for model training. Since our cognitive model consists of two parameters, it requires that the training data consists of the inter-keystroke timing sequences for at least two key pairs (1,350 key pairs are used in our experiment). The training data used in the learning phase can be collected from the attacker himself or people he/she recruits. The simplest way to collect training data is to implement a keylogger which records the key code of every keystroke event and the corresponding timestamp to get the ground truth.

\noindent\textbf{Cognitive Model Training.}
With the training data, the attacker can estimate the coefficients of the linear equation (Equation~\ref{formula}) in our cognitive model using the standard least squares method.

\noindent\textbf{Timing Dictionary Building.}
Once the cognitive model is fixed, the attacker can compute the inter-keystroke timing sequences for \textit{all} PINs and then generate a timing dictionary $D = \{(\text{PIN}_i, \overrightarrow{T_i})\}$ for $i=1, 2, ..., 10^l$ where $\overrightarrow{T_i} = (\Delta T_{i1}, \Delta T_{i2}, ... , \Delta T_{il})$ and $l$ is the PIN length. Here, $\Delta T_{ij}$ is computed according to the cognitive model for $j$-th key pair ($K_{ij}, K_{i(j+1)}$) in the $i$-th PIN ($K_{i1}, K_{i2}, ..., K_{il}$), where $j=1, 2, ..., l$ and $K_{i(l+1)} = <Enter>$. Table~\ref{timingdictionary} shows a segment in the inter-keystroke timing dictionary used in our experiments.

\subsection{Attacking Phase}
\label{attackingphase}
\noindent\textbf{Data Collection.}
In the attacking phase, an attacker needs to obtain a single inter-keystroke timing sequence $\overrightarrow{T}$ of a PIN entry made by a victim for PIN inference. Similar to each timing sequence in the timing dictionary, $\overrightarrow{T}$ is an $l$-dimensional sequence, where $l$ denotes the length of the target PIN.

\noindent\textbf{Similarity Calculation.}
Once the attacker has an observed timing sequence $\overrightarrow{T}$ of the target PIN (from a victim) and a timing dictionary $D$, he/she can measure the similarity between $\overrightarrow{T}$ and each timing sequence in $D$.

There are many similarity metrics the attacker can use. We test three different metrics (cosine similarity, Euclidean distance and Pearson product-moment correlation coefficient) and discover that the cosine similarity gives the best results in most attacks. The cosine similarity is a measurement of the level of similarity between two vectors $\overrightarrow{A}$ and $\overrightarrow{B}$ that returns the cosine of the angle between them and is computed as follows:
\begin{equation}
cos = \frac{\overrightarrow{A}\cdot\overrightarrow{B}}{\left\|\overrightarrow{A}\right\|\cdot\left\|\overrightarrow{B}\right\|}
=\frac{\sum_{i=1}^l{a_ib_i}}{\sqrt{\sum_{i=1}^l{a_i^2}}\cdot\sqrt{\sum_{i=1}^l{b_i^2}}},
\end{equation}
where $a_i$ and $b_i$ are the $i$-th elements of $l$-dimensional vectors $\overrightarrow{A}$ and $\overrightarrow{B}$, respectively. The time complexity for the similarity calculation is $O(n)$, where $n$ is the number of all possible PINs. The cosine similarity is scale-free, i.e., the amplitudes of $\overrightarrow{A}$ and $\overrightarrow{B}$ have no impact to the result. This feature improves the robustness of our attacks against variation of typing speeds between victims and different users in the training data, which thus contributes to the user independence of our approach.

\noindent\textbf{Ranking and using PINs in login attempts.}
The attacker then ranks all entries in the timing dictionary according to their similarity values so that those entries more similar to $\overrightarrow{T}$ appear closer to the top. Here, we use the Quicksort algorithm whose time complexity is $O(n*log(n))$ to rank all candidate PINs. Finally, the attacker attempts to login to the victim's account using the PINs starting from the top in the ranked list in an online attack.

\section{Experiments}
An IRB-approved user study is conducted to collect users' inter-keystroke timing data about PIN entries on a numeric keypad. The data collected from participants are kept confidential and anonymized. To examine the effectiveness of our attacks to different types of PINs, we study the inner structure of the whole PIN space and partition the PIN space into different strength levels. In this section, we present the performance of our attacks in the general attack setting in which the training data and testing data are collected from different users.

\begin{table} [!t]
	\centering
	\small
	\caption{List of PINs used in our experiments.}
	\begin{tabular}{|c|c|c|c|c|c|}
		\hline
		\multirow{2}{*} {\textit{Level 1}} & 777777 & 777333 & 222233 & 633333 & 555553 \\ \cline{2-6}
		                                   & 443333 & 088886 & 000553 & 055333 & 577773 \\ \hline
		\multirow{2}{*} {\textit{Level 2}} & 008853 & 166034 & 226633 & 515553 & 009666 \\ \cline{2-6}
		                                   & 800053 & 705333 & 100086 & 222253 & 100553 \\ \hline
		\multirow{2}{*} {\textit{Level 3}} & 911182 & 590253 & 537473 & 086483 & 084953 \\ \cline{2-6}
		                                   & 331086 & 410886 & 547733 & 537802 & 199993 \\ \hline
		\multirow{2}{*} {\textit{Level 4}} & 990872 & 098046 & 760973 & 301509 & 330117 \\ \cline{2-6}
		                                   & 301246 & 095653 & 589107 & 530271 & 603294 \\ \hline
		\multirow{2}{*} {\textit{Level 5}} & 420381 & 191061 & 806205 & 079039 & 033645 \\ \cline{2-6}
		                                   & 146928 & 501347 & 635210 & 684032 & 706759 \\ \hline
	\end{tabular}
	\label{pins}
\end{table}

\subsection{User Study}
Our user study involves 55 participants, including 24 males and 31 females with ages ranging from 19 to 34. All participants are students or members of staff at the Singapore Management University. Each participant is paid 10 Singapore dollars as a compensation for his/her time and effort. Since 6-digit PINs are commonly used in many PIN-based authentication systems, we use 6-digit PINs as examples of our attacks. Our user study consists of two sessions: training session and testing session. In both sessions, we use JavaScript to record the key code of each keystroke event and the corresponding timestamp to get the ground truth.

In the training session, 5 participants are asked to enter three 6-digit PINs (i.e., 146928, 501347, 635210) on a numeric keypad. The PINs they typed are randomly selected from the whole 6-digit PIN space. The participants are required to memorize one PIN intentionally, type the PIN for several times as exercises and type more times for data collection; then, they are required to forget the current PIN, and proceed in the experiment with the next PIN. In our experiments, we observe that exercises for five times are sufficient for a participant to type a 6-digit PIN fluently. Then the participants type each PIN for 15 times continuously for training data collection. We ensure that each PIN entry is typed correctly. If a participant enters incorrect digits and uses the $<$Delete$>$ or $<$Backspace$>$ key to correct an input, he/she is required to retype the PIN.

In the testing session, we choose 50 PINs with 10 PINs randomly selected from each of five PIN strength levels as listed in Table~\ref{pins}. The other 50 participants (except the five in training to make our attacks user independent) are asked to enter PINs on the same numeric keypad. Each participant is assigned to type 25 PINs with 5 PINs chosen randomly from the 10 PINs in each PIN strength level. Similar to the training session, the participants type each PIN for 5 times as practice and type each PIN for 15 times for testing data collection. In total, 225 PIN entries are collected for training and 18,750 PIN entries for testing.

The raw data of each PIN entry we collected consists of the timestamps of ($l$+1) keystroke events for $l$-digit PINs, where the last keystroke is for pressing the $<$Enter$>$ key. We define the inter-keystroke timing between keystrokes $K_i$ and $K_{i+1}$ as the difference between the two consecutive key-down times to cover both the time of finger movement between the two keys and the time for pressing the second key:
\begin{equation}
\Delta T_i = T_{K_{i+1}}^\downarrow - T_{K_i}^\downarrow.
\end{equation}
Therefore, the inter-keystroke timing sequence of each PIN entry that is used in our experiment is represented by an $l$-dimensional sequence $\overrightarrow{T} = (\Delta T_1, \Delta T_2, ... , \Delta T_l)$.

\subsection{PIN Strength Level}
We study the inner structure of the whole PIN space to examine the effectiveness of our attacks to different PINs. We propose an approach to partition the whole PIN space into different PIN strength levels according to the directional density of the inter-keystroke timing sequences in the timing dictionary. Each inter-keystroke timing sequence in the timing dictionary can be considered as an $l$-dimensional directional vector, where $l$ is the PIN length. Intuitively, if a PIN vector locates in a dense region according to the cosine similarity measurement in the vector space, it is more difficult for an attacker to single it out, that is, infer the PIN. This implies that such a PIN is more secure against our attacks since our attacks rank candidate PINs according to the cosine similarity between each entry in the timing dictionary and the observed timing sequence of a target PIN as explained in Section~\ref{attackingphase}. Based on this observation, we propose Algorithm~1 to measure the PIN strength of $l$-digit PINs.

\renewcommand{\algorithmicrequire}{\textbf{Input:}}
\renewcommand{\algorithmicensure}{\textbf{Output:}}
\begin{algorithm}
	\caption{\textbf{: PIN Strength Measurement}}
	\begin{algorithmic}[1]
		\Require A trained timing dictionary $D = \{(\text{PIN}_i, \overrightarrow{T_i})\}$ for $i=1, 2, ..., 10^l$ where $\overrightarrow{T_i} = (\Delta T_{i1}, \Delta T_{i2}, ... , \Delta T_{il})$.
		\Ensure The strength measurement $\overrightarrow{S_i}$ for each $\text{PIN}_i$.
		\For{each vector $\overrightarrow{T_i}$ in $D$}
		\State calculate the cosine similarity between $\overrightarrow{T_i}$ and all other vectors in $D$ and obtain a cosine similarity tuple $(cos_{i1}, cos_{i2}, .., cos_{i(i-1)}, cos_{i(i+1)}, .., cos_{i(10^l-1)})$ where $cos_{ij} = \frac{\overrightarrow{T_i}\cdot\overrightarrow{T_j}}{\left\|\overrightarrow{T_i}\right\|\cdot\left\|\overrightarrow{T_j}\right\|}$
		\State rank all cosine similarities in descending order and obtain a new tuple $(cos'_{i1}, cos'_{i2}, ..., cos'_{i(10^l-1)})$ where $cos'_{i1}\geq cos'_{i2}\geq ...\geq cos'_{i(10^l-1)}$
		\State $\overrightarrow{S_i} = (\overline{G}_1, \overline{G}_2, ..., \overline{G}_l)$ where $\overline{G}_j = \frac{1}{9*10^{j-1}} \sum_{10^{j-1}\leq n \leq 10^j-1} cos'_n$ and $j = 1, 2, ..., l$.
		\EndFor
	\end{algorithmic}
\end{algorithm}

Algorithm~1 takes a trained timing dictionary $D$ as the input. For each timing vector $\overrightarrow{T_i}$ for $\text{PIN}_i$ in $D$, the algorithm first calculates the cosine similarity between $\overrightarrow{T_i}$ and all other vectors in $D$. It then ranks all of the calculated cosine similarities in descending order and divide them into $l$ groups where the $j^{th}$ group consists of $(10^{j-1})^{th}$ to $(10^j-1)^{th}$ cosine similarities. Finally, it calculates the average value $\overline{G}_j$ of cosine similarities for group $j$, where $j = 1, 2, ..., l$. The algorithm output an $l$-dimensional tuple $\overrightarrow{S_i} = (\overline{G}_1, \overline{G}_2, ..., \overline{G}_l)$ to represent the PIN strength for each $\text{PIN}_i$, where $i=1, 2, ..., 10^l$. The overall time complexity of our PIN strength measurement algorithm is $O(l*10^{2l})$ and its space complexity is $O(l*10^{l})$.

With the strength measurement $(\overline{G}_1, \overline{G}_2, ..., \overline{G}_l)$ for all PINs, we partition the whole PIN space into ($l$-1) levels. First, an indirect stable sort with multiple keys is performed on all PINs. To be specific, it first ranks all PINs by key $\overline{G}_1$, if two PINs have the same value for key $\overline{G}_1$; then it ranks them by key $\overline{G}_2$; and so on. As a result, it ranks all $10^l$ PINs according to PIN strength in ascending order. The first 100 PINs after ranking are categorized into \textit{level 1} which includes the weakest PINs. The $101^{th}$ to $1000^{th}$ PINs are categorized into \textit{level 2}; the $1001^{th}$ to $10000^{th}$ PINs are categorized into \textit{level 3}; and so on. In our experiments, we take 6-digit PINs as examples and divide all 6-digit PINs into 5 categories. \textit{Level 1} to \textit{level 5} consist of 100, 900, 9,000, 90,000, 900,000 PINs, respectively.

\begin{figure} [!t]
	\begin{minipage}{\linewidth}
		\centering
		\includegraphics[width=0.8\linewidth]{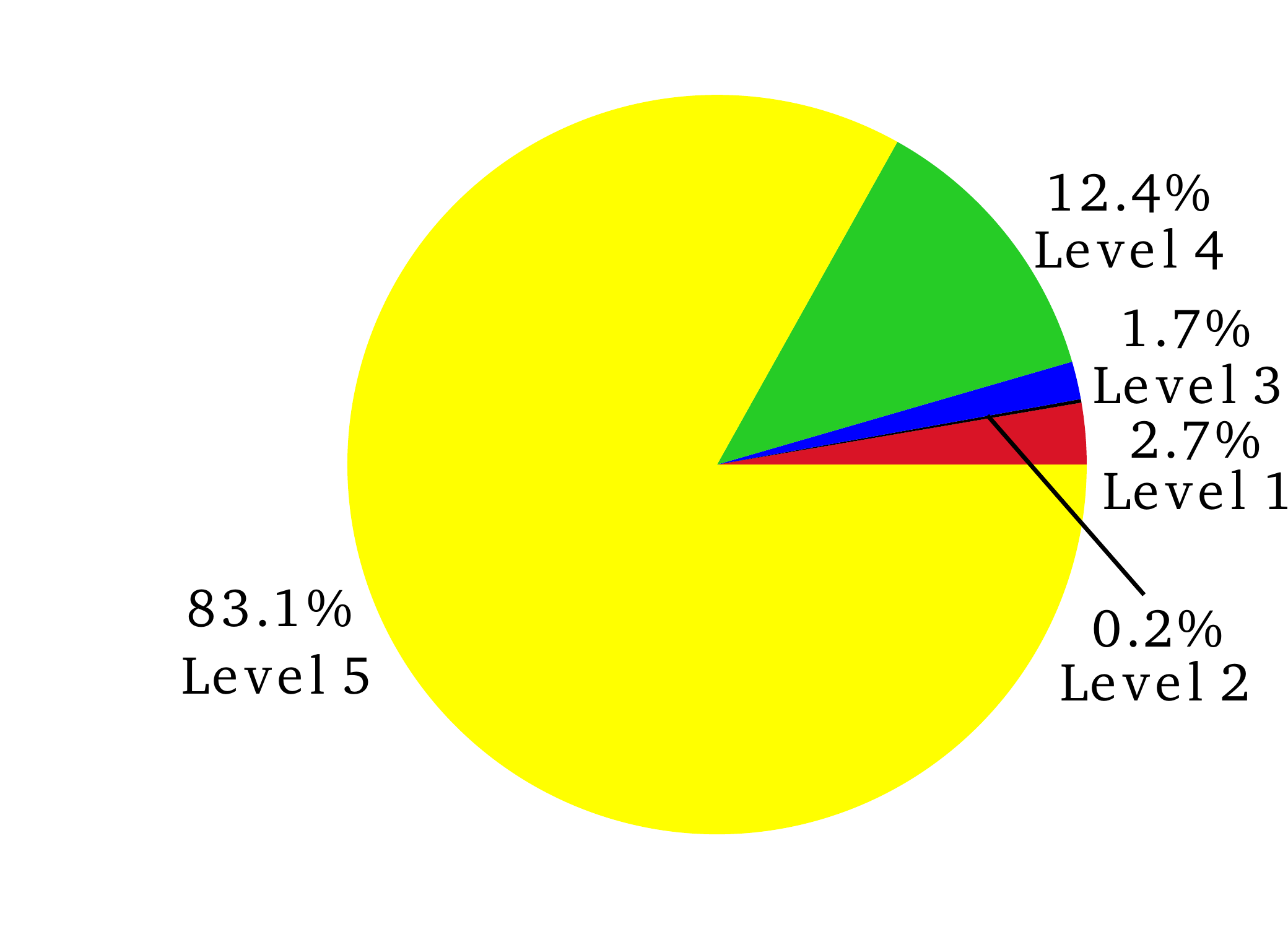}
		\caption{The proportion of human-chosen PINs at each strength level.}
		\label{fig:total}
	\end{minipage}
	\begin{minipage}{\linewidth}
		\centering
		\includegraphics[width=0.8\linewidth]{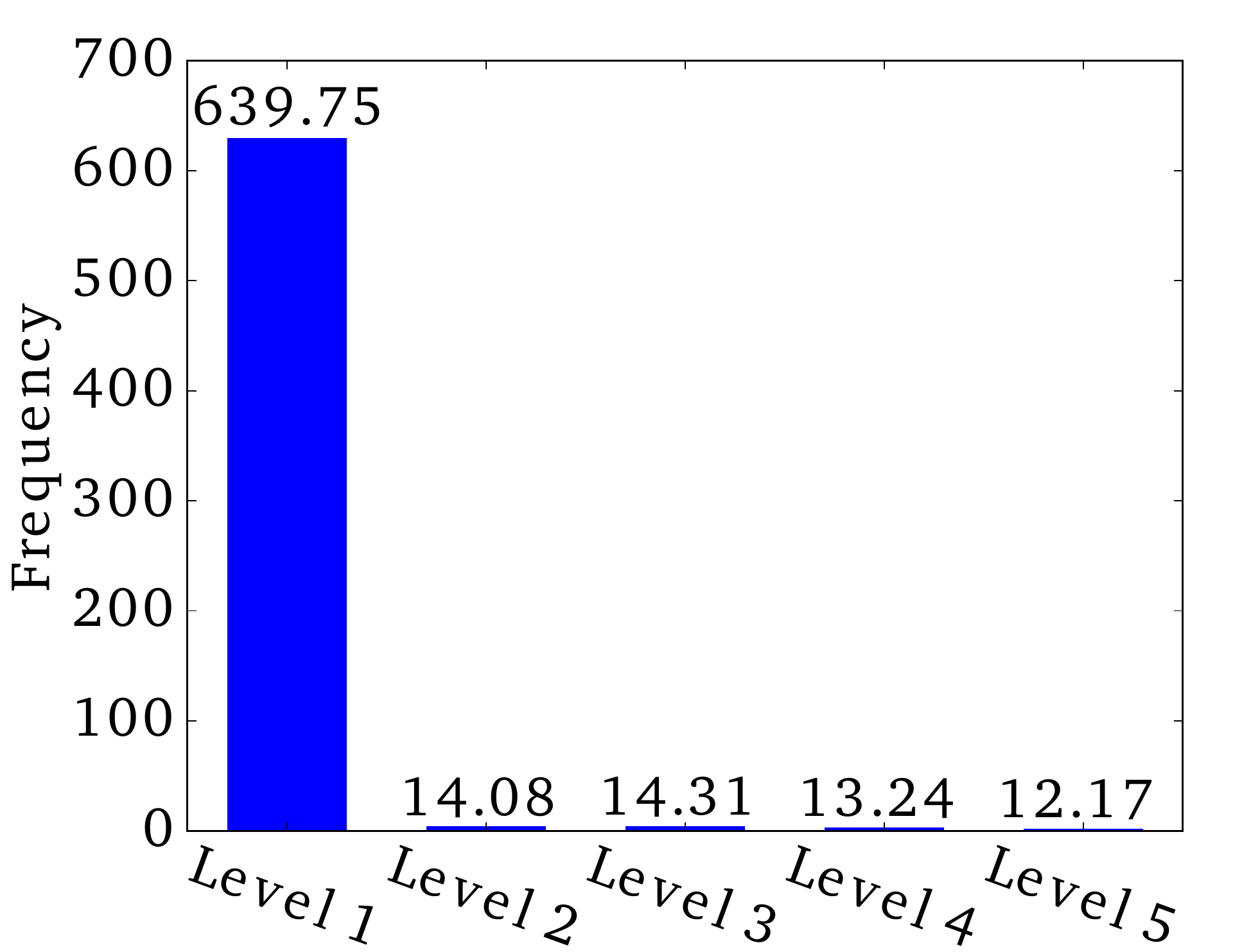}
		\caption{The averaged frequency of each PIN at different strength levels.}
		\label{fig:average}
	\end{minipage}
\end{figure}

We further study the distribution of human-chosen 6-digit PINs according to our PIN space partition. The human-chosen 6-digit PINs are extracted from two leaked large-scale password databases (i.e., Rockyou and CSDN). Figure~\ref{fig:total} shows the proportion of human-chosen PINs at each strength level and Figure~\ref{fig:average} shows the averaged frequency of each PIN at different strength levels. It is observed that although the PINs at the lowest security level (i.e., \textit{level 1}) account for only 0.01\% of the total, more than 2.7\% of real users prefer to select PINs at this level and the averaged frequency of at this level is significantly higher than other strength levels. These results show that users tend to select weak PINs more often than strong PINs. It is thus meaningful to evaluate PIN attacks at different security levels.

\subsection{Performance Evaluation}
\label{performance}
We evaluate the performance of our attacks in the general attack setting. First, we use the inter-keystroke timing data from the training session of our user study to train the cognitive model. The parameter $a$ and $b$ are 135.91 and 47.73, respectively. Based on this trained cognitive model, we estimate the timing sequence of all $10^6$ 6-digit PINs and generate a timing dictionary. According to our experiments, it takes 21.7s to generate a timing dictionary.

Then, we take each PIN entry typed by the participants from the testing session as an independent attacking case. In total, there are 18,750 individual cases for 50 PINs. Note that the training data and the testing data in our user study are collected from different groups of participants, which make our attacks user-independent. For each attacking case, we measure the cosine similarity between the observed timing sequence and each entry in the timing dictionary and rank all PINs according to their similarity values in the descending order. Given an observed timing sequence, it takes around 1s to get the ranking list of all candidate PINs in the general attack. If the correct PIN ranks $x$-th in the ranked list, an attacker needs to login to the target victim's account for $x$ times until success.

\begin{figure} [!t]
	\centering
	\includegraphics[width=\linewidth]{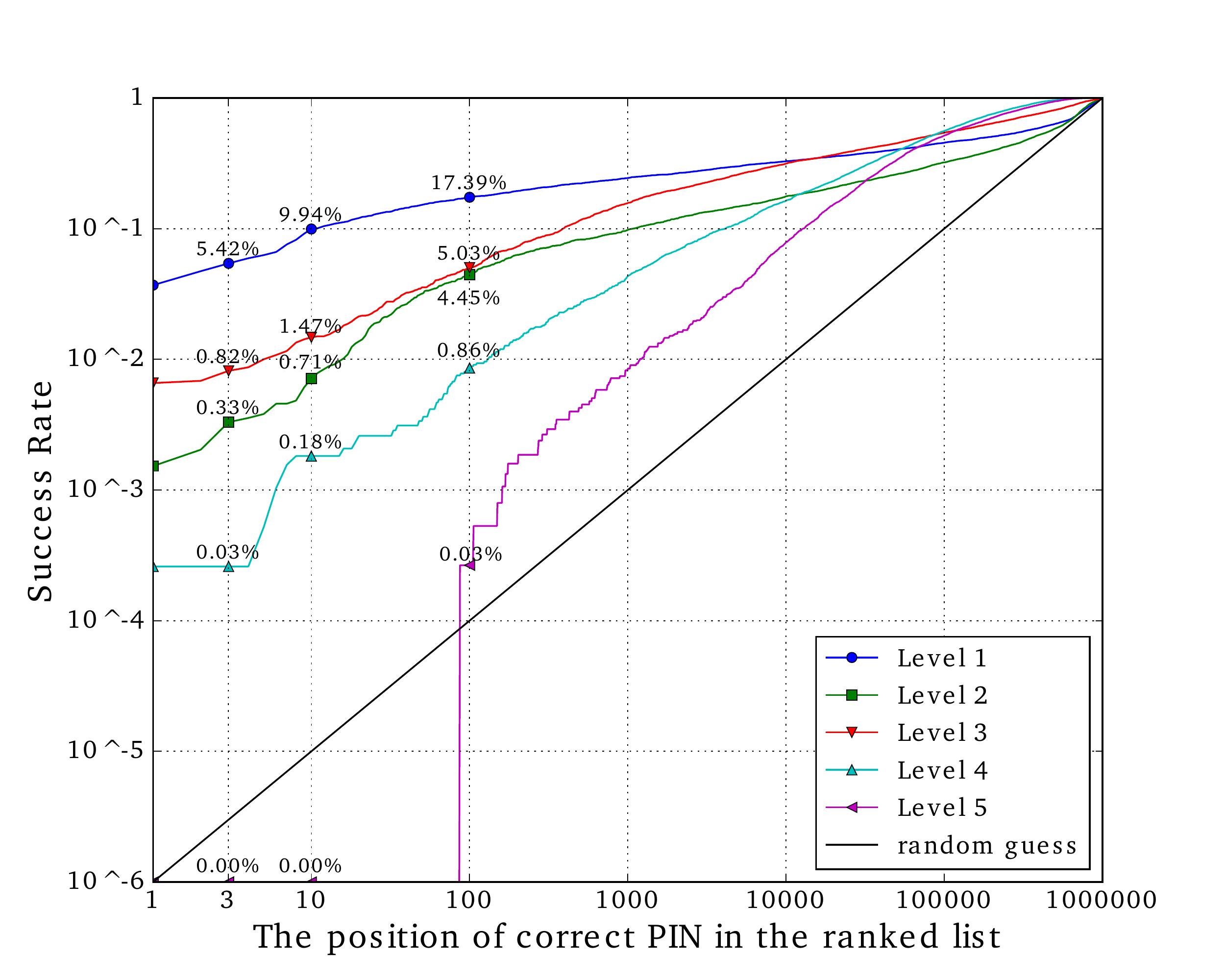}
	\caption{The performance of \textit{general attacks}.}
	\label{fig:gen}
\end{figure}

The performance of such \textit{general attacks} is shown in Figure~\ref{fig:gen}, where the x-axis denotes the position of a correct PIN in the ranked list and the y-axis denotes the success rate of hitting the correct PIN in an attack. The success rate of our attacks is calculated as the observed frequency that the correct PIN appears in the top $x$ ranked PINs across all attacking cases. Note that the success rate of our attacks is 0 before any successful case is observed. Figure~\ref{fig:gen} also shows the success rate of random guessing attacks, assuming that the correct PIN has an equal probability to appear at any position between 1 and $10^6$. The success rate of random guessing attacks is $\binom{10^l-1}{x-1}/\binom{10^l}{x}$ for an $l$-digit PIN where $x$ is the maximum number of allowed consecutive failures.

A general trend in Figure~\ref{fig:gen} is that it is more effective to attack PINs at lower strength levels. Beyond our expectation, the performance of PINs at \textit{level 3} is better than \textit{level 2} but the difference between them is not too significant. Maybe it is because that number of samples in each levels is small in our user study. This trend suggests that users should choose their PINs at the strongest strength level for better security in the presence of inter-keystroke timing attacks.

Another trend in Figure~\ref{fig:gen} is that the performance of \textit{general attacks} is much better than random guessing attacks. In particular, if the number of allowed attempts is limited to 100, 10 and 3, our \textit{general attacks} improve the success rate by 522, 2247 and 4004 times on average of all PIN strength levels over the random guessing attacks, respectively.

Our experimental results imply that the existing PIN-based authentication systems are vulnerable to our attacks, especially when they are launched at a large scale. When a victim types a PIN at \textit{level 1}, an attacker can launch a successful attack within 10 consecutive attempts with a probability about 10\%. It has been argued that if 10\% of accounts in an authentication system are compromised, an attacker may access all resources of the system~\cite{florencio2016pushing}. 

\section{Other Specific Attacks}
\label{otherattacks}
While the \textit{general attacks} we discussed in the previous section are user-independent (i.e., the training data and the testing data are collected from different users), we examine other specific attacks to improve the success rate of \textit{general attacks} with different assumptions on attackers' capabilities.

\subsection{Target Attacks}
We first examine whether the performance of our attacks can be improved using target victims' data for model training which is also used in HMM-based attacks in the literature~\cite{song2001timing, zhang2009peeping, KuneKim2010TimingPIN}. Hence, we propose \textit{targeted attacks} where an attacker obtains a small amount of inter-keystroke timing data about a victim typing known numerical sequences for model training. Although both \textit{targeted attacks} and HMM-based attacks train their models based on a victim's own data, our approach requires much less training data. Our approach requires an attacker to know the inter-keystroke timing data about a few key pairs rather than all key pairs as required in HMM-based attacks. To collect such training data in practice, an attacker may trick a victim to install malware on his/her smartphone and collect inter-keystroke timing data when the victim dials phone numbers. Another possible way of collecting such data is to trick a victim to enter insensitive numerical sequences through phishing websites or phishing phone calls.

The procedure of the experiment of \textit{targeted attacks} is similar to that of \textit{general attacks} except that we use the inter-keystroke timing data from the testing session of our user study to train a cognitive model. In particular, to attack any one of the 25 PINs entered by a participant, we randomly choose 2 other PINs out of the 25 PINs entered by the same participant and use 30 collected inter-keystroke timing sequences for these 2 PINs for model training. In comparison, previous HMM-based attacks require that an attacker should obtain 30-50 inter-keystroke timing sequence for \textit{each of 110 key pairs} ($10\times10$ digit-to-digit key pairs and 10 digit-to-$<$Enter$>$ key pairs) from a victim for model training. The same as the \textit{general attacks}, we take each PIN entry typed by the participants from the testing session as an independent attacking case.

\begin{figure} [!t]
	\centering
	\includegraphics[width=\linewidth]{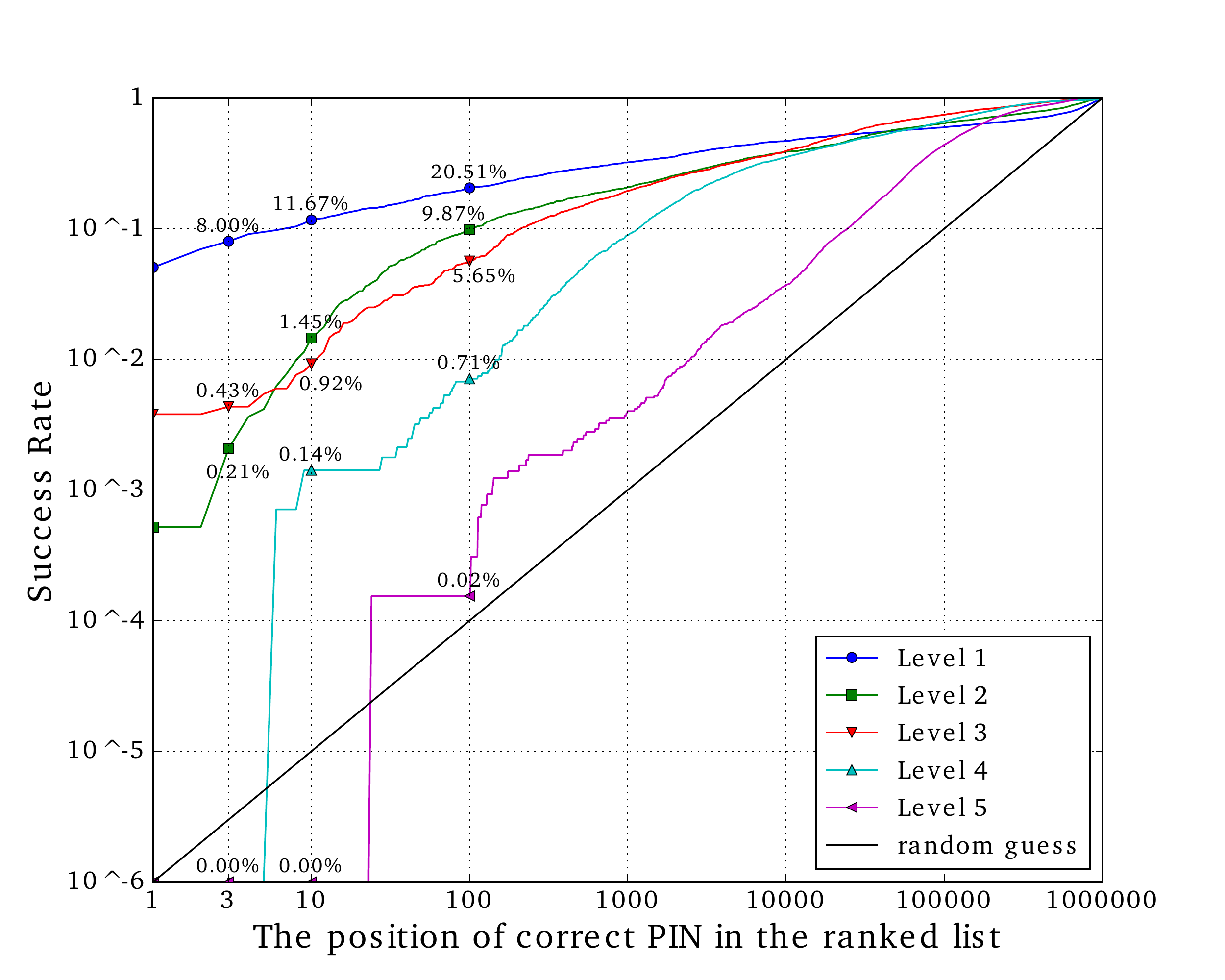}
	\caption{The performance of \textit{targeted attacks}.}
	\label{fig:tar}
\end{figure}

Figure~\ref{fig:tar} shows that \textit{targeted attacks} have a similar trend as \textit{general attacks} in terms of the effectiveness of attacking PINs at different PIN strength levels. Compared to \textit{general attacks}, the success rate of \textit{targeted attacks} is improved by about 4\% on average for all levels. Considering that \textit{targeted attacks} are user dependent, and they do not improve the success rate significantly over the \textit{general attacks}, attackers may still prefer \textit{general attacks} in practice.

\subsection{Multi-Entry Attacks}
\begin{figure} [!t]
	\centering
	\includegraphics[width=\linewidth]{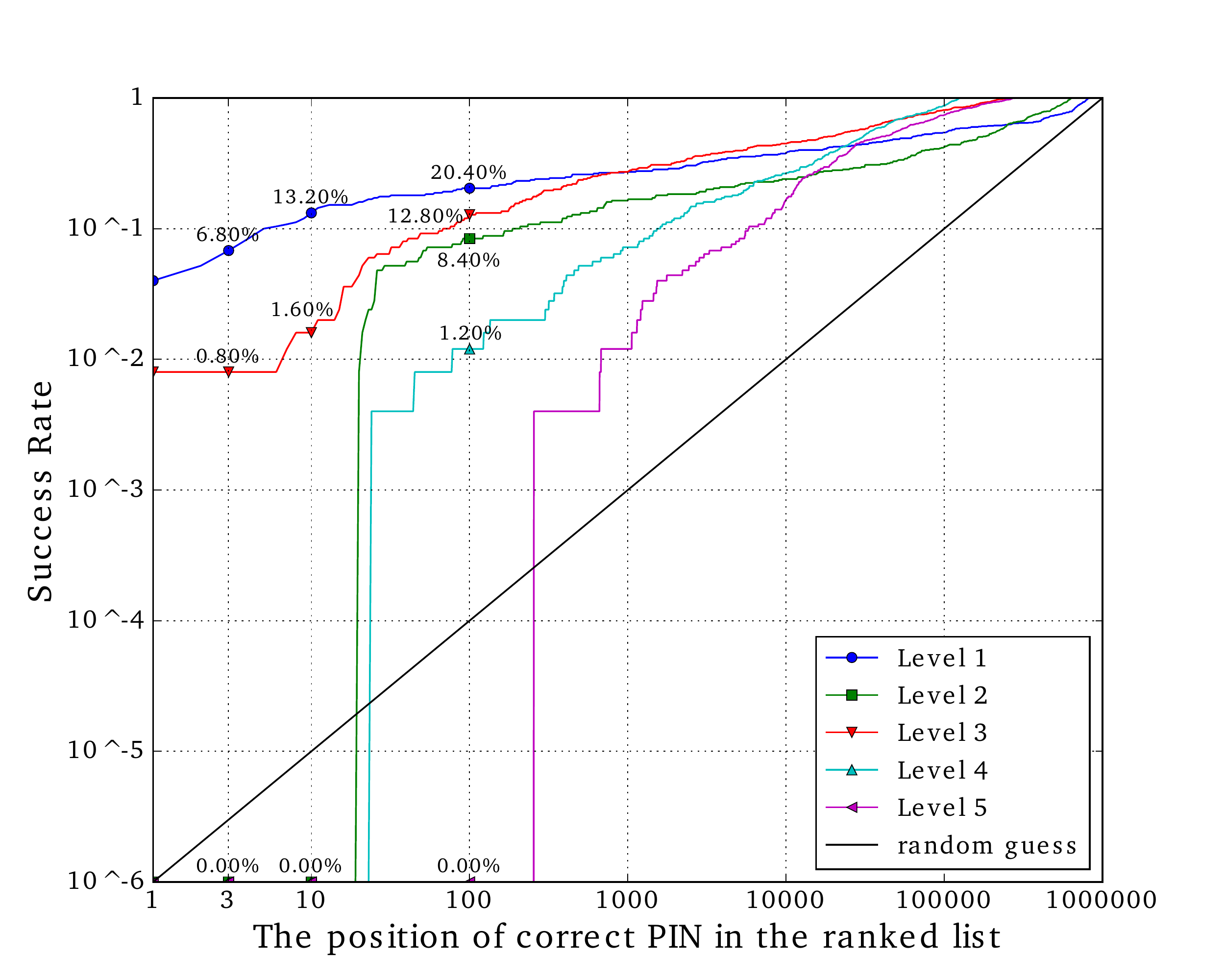}
	\caption{The performance of \textit{multi-entry attacks}.}
	\label{fig:mt}
\end{figure}
We then examine whether an attacker can improve his/her success rate if he/she observes the victim's PIN entry for multiple times. Hence, we propose \textit{multi-entry attacks} where an attacker captures the inter-keystroke timing sequences about a victim entering the same PIN for multiple times. With $k$ inter-keystroke timing sequences of one PIN, an attacker can calculate an averaged timing sequence for PIN inference.

First, the attacker normalizes each observed PIN entry's inter-keystroke timing sequence so as to attain the same amplitude. The ratio of $\text{Sum}_i$ to $\overline{\text{Sum}}$ is considered as the scaling value for the $i$-th inter-keystroke timing sequence $\overrightarrow{T_i} = (\Delta T_{i1}, \Delta T_{i2}, ..., \Delta T_{il})$, where $\text{Sum}_i = \sum_{1\leq j \leq l} \Delta T_{ij}$, $\overline{\text{Sum}} = \frac{1}{k} \sum_{1\leq i \leq k} \text{Sum}_i$, and $l$ is the PIN length.

Then, the attacker calculates the $i$-th scaled inter-keystroke timing sequence $\overrightarrow{T'_i} = \overrightarrow{T_i} \times (\overline{\text{Sum}}/\text{Sum}_i)$. Given $k$ scaled timing sequences $\overrightarrow{T'_1} = (\Delta T'_{11}, \Delta T'_{12}, ..., \Delta T'_{1l})$, ..., $\overrightarrow{T'_k} = (\Delta T'_{k1}, \Delta T'_{k2}, ...,\\ \Delta T'_{kl})$, the attacker generates an averaged timing sequence $(\Delta \overline{T'_1},\\ \Delta \overline{T'_2}, ..., \Delta \overline{T'_l})$ where $\Delta \overline{T'_j} = \frac{1}{k} \sum_{1\leq i \leq k} \Delta T'_{ij}$.

Similar to the \textit{general attacks}, the attacker trains a cognitive model from other users' inputs and builds a timing dictionary. The attacker then calculates the similarity between the calculated averaged timing sequence and each entry in the timing dictionary and ranks all PINs according to their similarity values. Finally, the attacker attempts to login to the victim's account using the PINs starting from the top in the ranked list in an online attack.

In the experiment of \textit{multi-entry attacks}, the same cognitive model and timing dictionary are used as in \textit{general attacks}. We take the averaged inter-keystroke timing information of 10 PIN entries (i.e., $k=10$) from each participant as an independent case. The procedure of the experiment of \textit{multi-entry attacks} is similar to that of \textit{general attacks} except that we take the averaged timing sequence as the observed timing sequence in each attacking case.

Figure~\ref{fig:mt} shows that \textit{multi-entry attacks} have a similar trend as \textit{general attacks} in terms of the effectiveness of attacking PINs at different PIN strength levels. Compared to \textit{general attacks}, \textit{multi-entry attacks} achieves better performance when $x$ ranges from $100$ to $10^6$ but achieves worse performance to the PINs at \textit{level 2, 4, 5} when $x$ ranges from $1$ to $100$. One possible reason is that the number of samples in \textit{multi-entry attacks} is much less than \textit{general attacks} and the observation of finding the position of correct PIN in the top 100 is based on a large number of samples. In general, \textit{multi-entry attacks} outperform \textit{general attacks} with insignificant improvement (below 2\% on average for all levels).

\subsection{Known Digits Attacks}
\begin{figure} [!t]
	\centering
	\includegraphics[width=\linewidth]{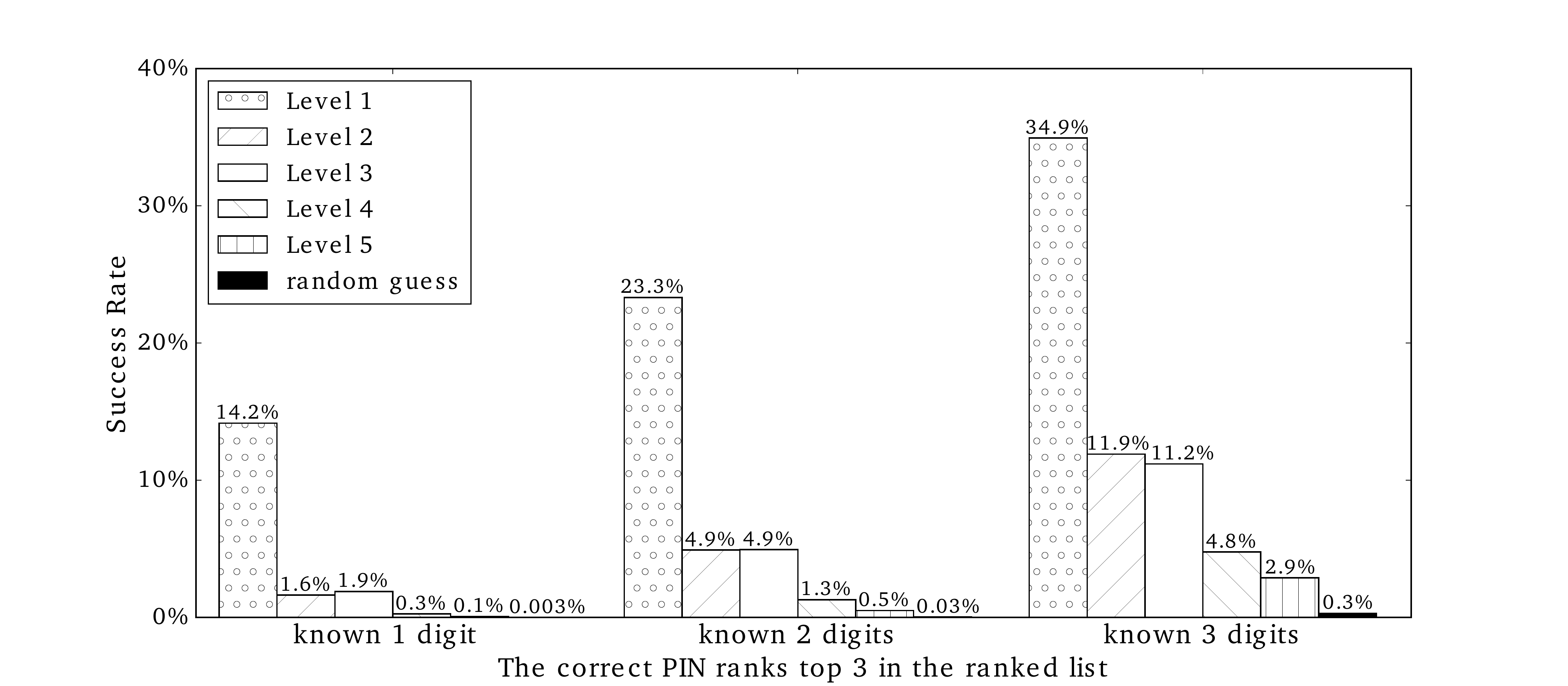}
	\includegraphics[width=\linewidth]{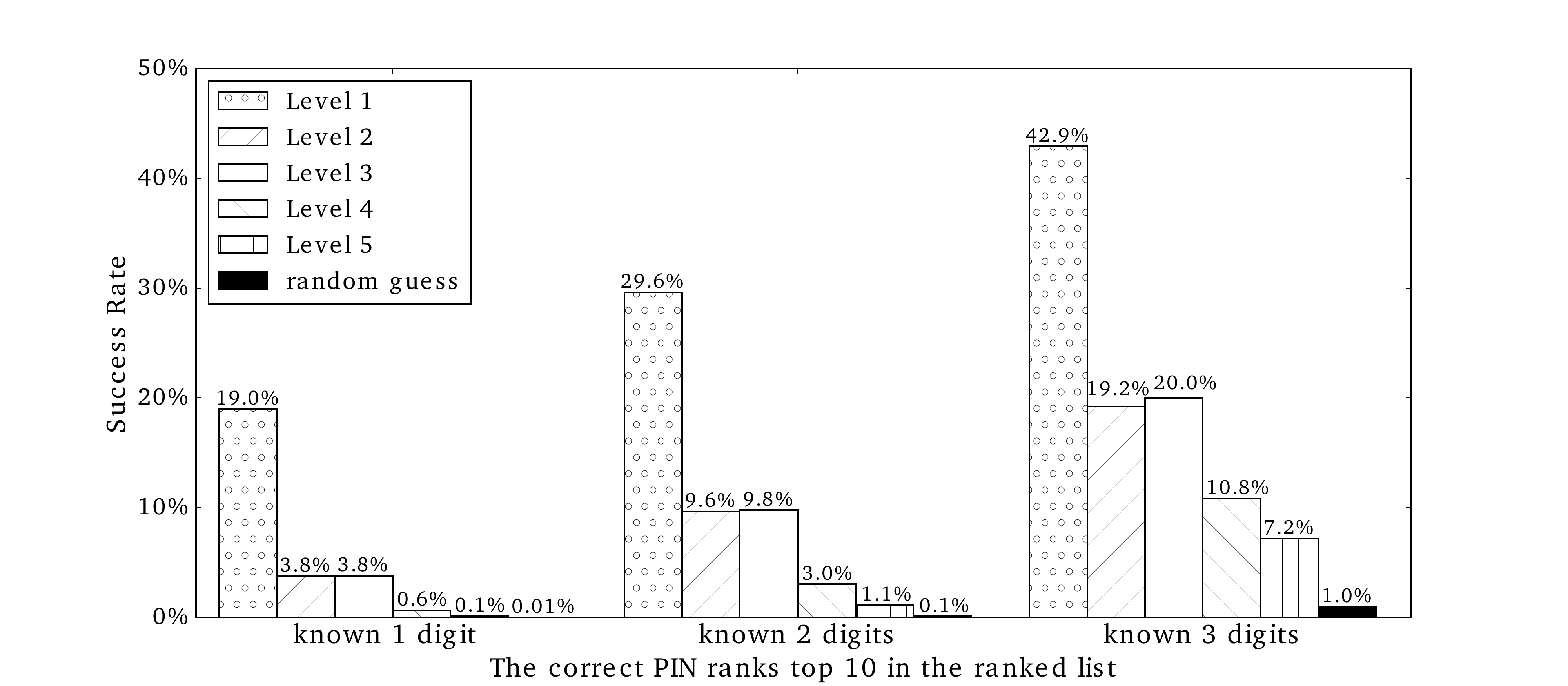}
	\includegraphics[width=\linewidth]{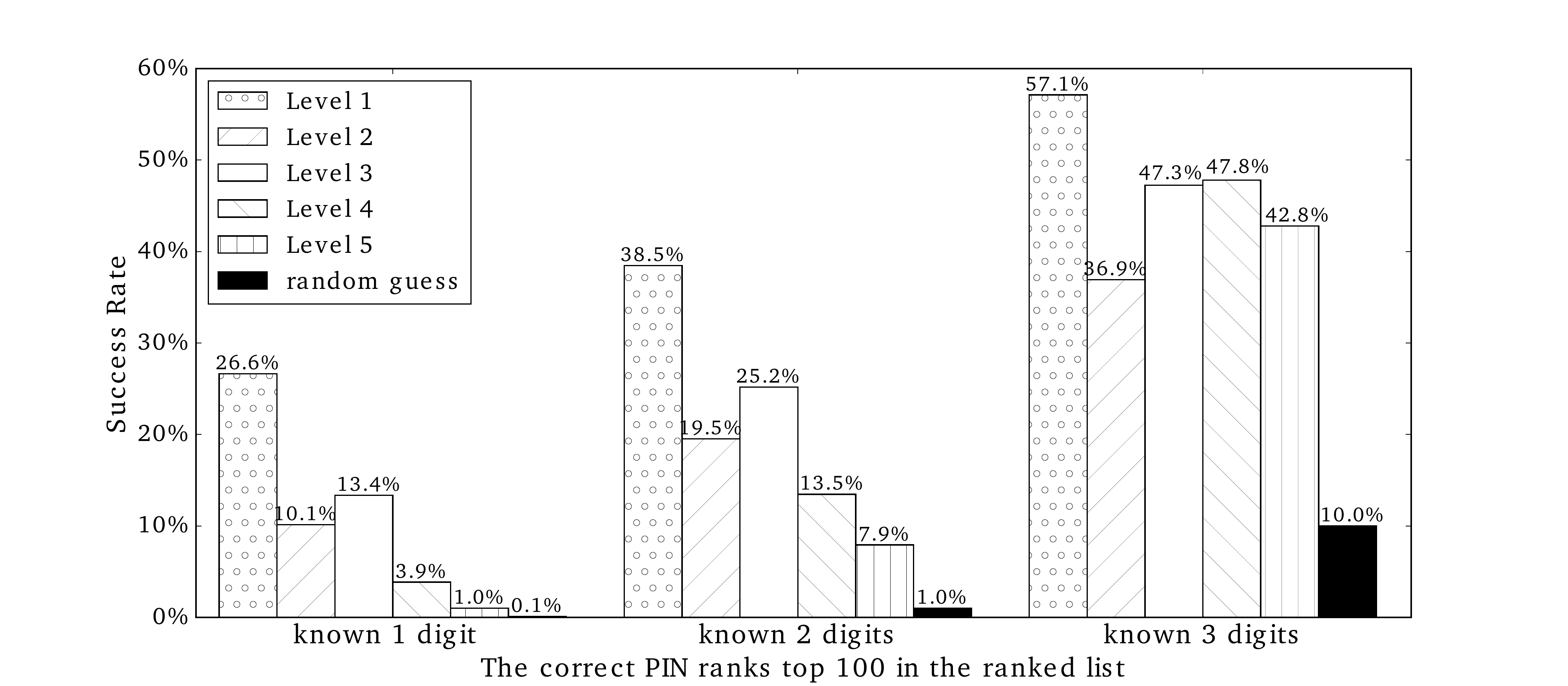}
	\caption{The performance of \textit{known digits attacks}.}
	\label{fig:kln}
\end{figure}
Considering that both \textit{targeted attacks} and \textit{multi-entry attacks} bring little improvement over \textit{general attacks}, we further propose \textit{known digits attacks} which improve the success rate significantly.

In this case, an attacker knows certain digits of a target PIN before launching inter-keystroke timing attacks (e.g., through other side channel attacks\cite{de2016identification, yue2014blind, shukla2014beware, liu2015good, wang2016Friend, sun2016visible} or shoulder surfing attacks~\cite{tari2006comparison}). Hence, he/she can reduce the size of his/her timing dictionary. For example, if the first two digits are known to the attacker which are `1' and `2', the reduced timing dictionary consists of $10^4$ candidate PINs which range from `120000' to `129999'. The attacker measures the similarity between the observed timing sequence and each timing sequences in the reduced timing dictionary and ranks these $10^4$ candidate PINs according to their similarity values. Finally, the attacker attempts to login to the victim's account using the PINs starting from the top in the ranked list in an online attack.

In the experiment of \textit{known digits attacks}, we use the same cognitive model as in \textit{general attacks} and generate reduced timing dictionaries. We evaluate the cases where an attacker obtains 1, 2 or 3 digit(s) of a target PIN. For each attacking case, one inter-keystroke timing sequence for each PIN entry is used. We enumerate all cases where the known value(s) are at any position(s) of the target PIN (i.e., 6 cases for known 1 digit, 15 cases for known 2 digits and 20 cases for known 3 digits for each PIN entry). For the similarity calculation, we measure the similarity between the observed timing sequence and each entry in the corresponding reduced timing dictionary. When an attacker knows any $k$ digits of an $l$-digit PIN, the success rate of random guessing attacks is $\binom{10^{l-k}-1}{x-1}/\binom{10^{l-k}}{x}$ where $x$ is the maximum number of allowed consecutive failures.

The results of \textit{known digits attacks} are shown in Figure~\ref{fig:kln}. It is clear that the success rate of \textit{known digits attacks} is significantly higher than \textit{general attacks}. For example, the success rates of inferring a target PIN at \textit{level 1} within 3 attempts are 14.2\%, 23.3\%, and 34.9\% when 1 digit, 2 digits, and 3 digits are known by the attacker, and they are 1.6, 3.3, 5.4 times higher than \textit{general attacks}, respectively. In many cases, the success rate of guessing the correct PIN is above 10\%. The results show that \textit{known digits attacks} are more practical than \textit{general attacks}. Even \textit{known digits attacks} are applied to attack a single user or a small number of users, their success rates are not impractically low. These results also indicate the effectiveness of our inter-keystroke timing attacks to 3, 4 or 5-digits PINs and that the attacks pose a greater threat to shorter PINs as expected.

\section{Discussions}
In this section, we compare our attacks with HMM-based attacks to show the merits of our approach. Then, we demonstrate that our attacks pose a serious threat to real-world applications when applied at large scale. Next, we propose several feasible countermeasures to mitigate our attacks. Lastly, we discuss the limitations of our attacks.

\subsection{Comparison with HMM-based Attacks}
Most keystroke timing attacks in the literature follow a similar attacking framework based on a Hidden Markov Model (HMM)~\cite{song2001timing, zhang2009peeping, KuneKim2010TimingPIN}. Compared to HMM-based attacks, our attacks have two merits.

The first merit is that our general attacks are user-independent. The cognitive model in our attacks captures the common characteristics across all skilled users typing PINs so that it can be used to attack any users. In addition, the use of cosine similarity in our attacks enables an attacker to rank all candidate PINs similarly for inferring a target PIN even if different users may type the target PIN with different speeds. In comparison, the HMM-based attacks relies on the distribution of inter-keystroke timing for a specific user typing each possible key pair so as to calculate the probability of any possible underlying keystroke sequence given an observed inter-keystroke timing sequence. Because the distribution of inter-keystroke timing for different users typing any same key pair may not be similar, the HMM-based attacks are user-dependent. They require that an HMM be trained with the inter-keystroke timing data for all possible key pairs collected from a target user, and such a model is user dependent and has to be retrained from scratch if the target user changes.

The second merit is that the cognitive model used in our attacks can be trained based on inter-keystroke time intervals for a small number of key pairs (minimum two key pairs). To launch an HMM-based attack, however, an attacker needs to collect a sufficiently large number of inter-keystroke time intervals for \textit{each possible} key pair from a target user before launching the attack. For PIN inference, an attacker needs to capture 30-50 inter-keystroke time intervals for each of 110 key pairs (including $10\times10$ digit-to-digit key pairs and 10 digit-to-$<$Enter$>$ key pairs) from a target user. It is usually difficult for an attacker to collect such large amount of data before launching an online attack in practical settings. Under the adversary model of our attacks, attackers cannot collect enough training data to support HMM-based attacks. 

\subsection{Attack Threats to Real-World Applications}
In general, the success rate of the proposed attacks may not be sufficiently high to pose imminent danger to an individual user's PIN if the attacker does not have prior knowledge on any digits of the target PIN. However, our attacks are practical in online settings because the attacks are user-independent and thus can be applied to attack any number of users' PINs in a large scale. To show the threats of our attacks to real-world applications, we provide two examples where PINs are used as the only credential to protect users' accounts and where attackers can collect many users' inter-keystroke timing data for PIN entries using malicious JavaScripts.

One example is the internet banking system of bank with pseudonym XYZ, which is the largest bank in a Southeast Asia country. It has more than three million Internet banking users. To login to an Internet banking account, a user needs to input a user ID and a 6 to 9-digit PIN as the credential (most users choose 6-digit PINs, which is the default case). Our tests show that users are not blocked within 50 login attempts. Although certain financial services (e.g., bank transactions) require a second-factor authentication, much sensitive information (e.g., account balances, usernames, addresses) can be leaked merely after PIN authentication. If ten percent of users' inter-keystroke timing data about PIN entries were collected, our online attacks can be applied to all these users' accounts with 50 tries per account, which do not lead to any account being locked in practice. Consequently, On the average, around 4.16\% of users' accounts would be compromised according to our experimental results. In other words, more than 12,000 users' accounts would be compromised due to our attacks.

The other example is ABCpay (pseudonym) which is the largest third-party mobile and online payment platform in Asia. It has more than 520 million users over the world. To make a payment through its service, a user needs to input his/her mobile phone number as user ID and a 6-digit PIN as password. It is not difficult for attackers to obtain many users' names, mobile phone numbers, and email addresses by crawling public web pages. The login attempts of each user's account in this platform is limited to 3. On the average, an attacker needs to launch online attacks to 83 users' accounts in order to compromise one account. In other words, if our attacks were applied to 1/1000 of users' accounts, then 6,000 users' accounts would be compromised on the average. Our attacks would cause serious damages in this case since attackers can transfer money from victims' accounts to other accounts.

Considering that many financial institutions have a large number of users and that malicious JavaScripts are easy to spread, our attacks pose a serious threat to real-world applications when applied in large scale.

\subsection{Mitigations}
\label{mitigation}
\noindent\textbf{Increasing PIN length.} The security strength of the most existing PIN systems is chosen according to the success probability of random guessing attacks~\cite{Yan2012LRPS}. For example, the security strength for 6-digit PINs is considered to be $10^{-6}$. However, our study reveals that the inter-keystroke timing attacks significantly lower the security strength of PIN systems. A simple approach to mitigating this threat is to increase the PIN length. Our calculation suggests that users should increase 6-digit PINs to 10-digit PINs whose security strength under the inter-keystroke timing attacks is higher than that of 6-digit PINs under the random guessing attacks on the average. This mitigation does not require any change to the hardware of current PIN authentication systems, but at the expense of requiring users to memorize longer PINs.

\noindent\textbf{PIN selection policy.}
As shown in Section~\ref{performance}, the performance of our attacks varies significantly when they apply to PINs at different strength levels. If a user selects a 6-digit PIN at \textit{level 5} instead of \textit{level 1} to \textit{level 4}, the probability of a successful \textit{general attack} within 100 attempts can be reduced by 870, 222, 251, and 43 times, respectively.

We thus suggest adopting a PIN selection policy where a user is required to choose a PIN at \textit{level 5} when the user registers his/her account. If a user chooses a PIN at \textit{level 1} to \textit{level 4}, his/her registration would not succeed until the user changes his/her PIN to \textit{level 5}. \textit{Level 5} consists of $9*10^5$ PINs which account for 90\% of all 6-digit PINs. It is thus relatively easy for a user to obtain a PIN at \textit{level 5} if he/she simply chooses his/her PIN randomly.

Considering that the success rate of attacking 6-digit PINs at \textit{level 5} is still around 10 times higher than random guessing attacks, To achieve a similar security strength of 6-digit PINs under the random guessing attacks, we suggest users choose 7-digit PINs at the strongest strength level. Note that when the PIN selection policy is adopted, it is unnecessary for users to choose 10-digit PINs which has been mentioned earlier.

\begin{figure} [!t]
	\centering
	\includegraphics[width=0.6\linewidth]{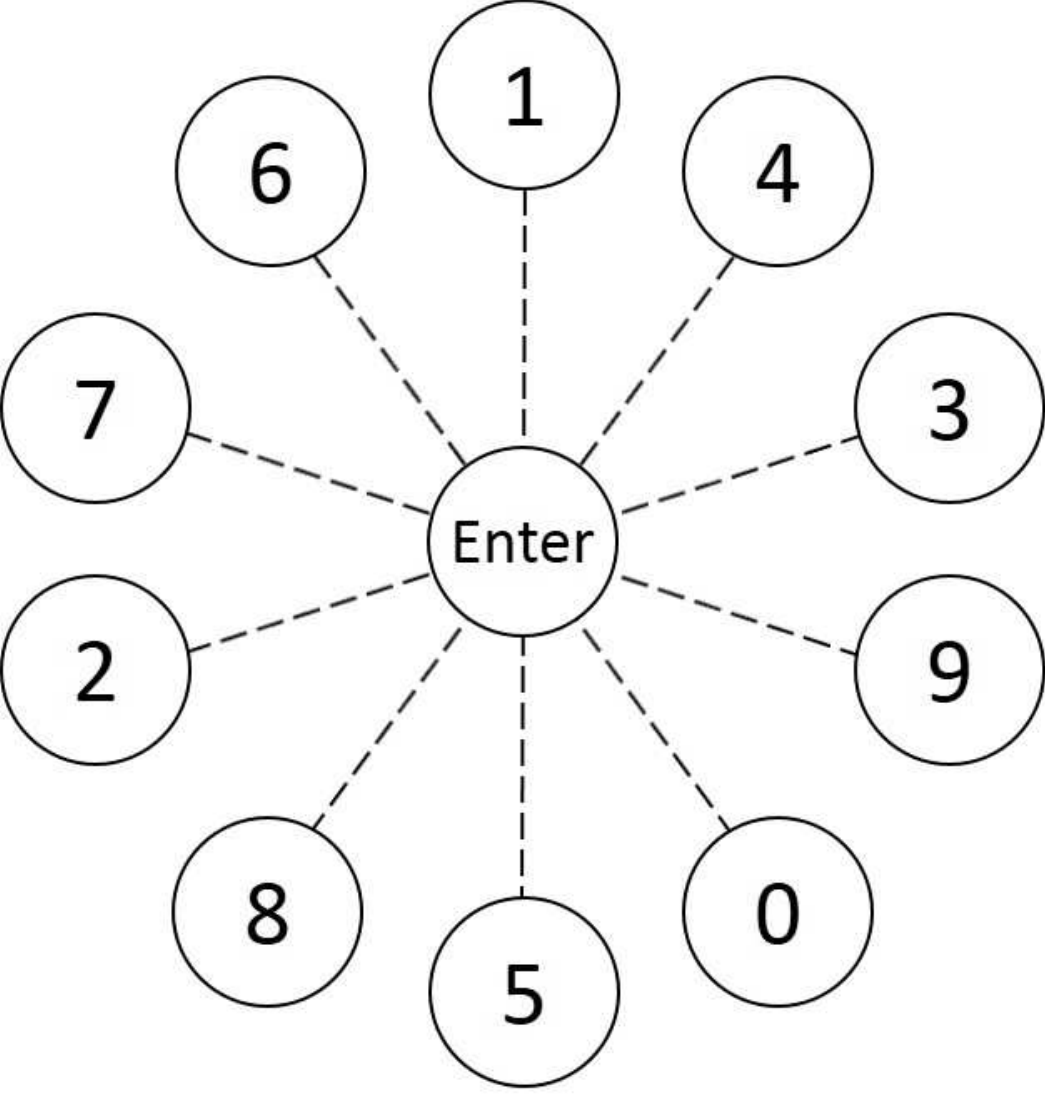}
	\caption{A new keypad layout.}
	\label{fig:newkeypad}
\end{figure}
\noindent\textbf{A new keypad layout.} As it is shown in the cognitive model, the inter-keystroke timing measurement for a user types a key pair on a keypad is mainly determined by the distance between the two keys of the key pair on the keypad. Based on this observation, we design a new keypad for PIN entry to mitigate inter-keystroke timing attacks. As shown in Figure \ref{fig:newkeypad}, the keypad is in a circular shape. All 10 digits (0-9) keys are evenly distributed on a circle. An $<$Enter$>$ key is located in the center of the keypad. When a user types his/her PIN, the user presses $<$Enter$>$ key after pressing each digit. When submitting the PIN, the user may press the $<$Enter$>$ key twice. During the PIN entry, the user always moves his/her finger through the same distance for entering any digit, leading to similar inter-keystroke timing sequence for entering any PINs.

Although a user may take double time for entering his/her PIN on this new keypad, the security strength of a PIN system is improved significantly against the inter-keystroke timing attacks. If this new keypad is adopted, the success rate of inter-keystroke timing attacks would be similar to that of random guessing attacks. We implement this keypad on a smartphone where the distances between any digit key and $<$Enter$>$ key is 1 inch. It takes around 2.5 seconds for a user to enter a 6-digit PIN on the new keypad. In comparison, most existing leakage resilient password systems which have the same security strength as that of 6-digit PINs require hundreds of seconds for user authentication (e.g.~\cite{hopper2001secure, asghar2010new, weinshall2006cognitive, li2005secure}).

\textbf{Leakage resilient password systems (LRPSes).} LRPSes~\cite{asghar2013does, hopper2001secure, asghar2010new, weinshall2006cognitive, li2005secure, bai2008pas, wiedenbeck2006design, yan2013designing} are user authentication systems which do not disclose user credentials to observers. Such systems are by design secure against any side-channel attacks including our attacks. However, Yan et al.~\cite{Yan2012LRPS} point out that in order to be secure, LRPSes have remarkably low usability. A secure LRPS usually takes hundreds of seconds to complete an authentication session, which may not be practical in many applications~\cite{hopper2001secure, asghar2010new, weinshall2006cognitive, li2005secure}.

\subsection{Limitations}
\label{limitation}
\noindent\textbf{Ecological validity}. In our user study, we recruit students and young staff only from a single university. The performance of our attacks may vary among different populations. The ecological validity of our user study is limited, but the qualitative facts in our research are likely to remain true.

\noindent\textbf{Typing styles}. In our experiments, we require all participants to enter their PINs on a keypad using a single finger. We conducted a larger scale survey on user's typing habits through emails and social networks over three weeks. In total, we received 544 responses. The participants of the survey mainly came from Singapore, China and the UK. They were not limited to the students or staffs in universities. According to the survey results, most participants reported that they tend to use a single finger when they enter PINs on numeric keypads in real life. In particular, 344 participants (63.2\%) use one finger, 124 participants (22.8\%) use two fingers, and 76 participants (14.0\%) use more than two fingers for PIN entry. In order to attack users who use multiple fingers when typing PINs, our cognitive model should be extended to cover different typing styles.

\noindent\textbf{Typing error}. During the process of entering PINs in real-world scenarios, users may press a wrong digit and then use the $<$Delete$>$ or $<$Backspace$>$ key to cancel the wrong input. We exclude this situation because it rarely happens in PIN entries. We plan to address this issue and generalize our attacks for password inference in the future.

\section{Related Work}
Besides inter-keystroke timing attacks, many other side-channel attacks for keystroke inference have been proposed in the literature recently. We summarize three types of side-channel attacks, including audio-based attacks, video-based attacks, and sensor-based attacks.

The first type of side-channel attacks for keystroke inference is audio-based attack. Asonov et al.~\cite{asonov2004keyboard} and Zhuang et al.~\cite{zhuang2009keyboard} demonstrate that the sounds of hitting different keys are different and thus propose machine learning algorithms to classify them. However, their approaches are user-dependent and keyboard-dependent. In comparison, the cognitive model in our attacks can be built for attacking any users as long as the geometric measurements of victims' keypads are known by attackers.

Different from machine learning based approaches, a training-free method is proposed by Berger et al.~\cite{berger2006dictionary}. They observe that the similarity of two keystroke acoustic signals has a negative relationship with the distance between the corresponding two keys. They infer English words typed by a user according to the relationship patterns of keystroke acoustic signals of all words in an English dictionary. In contrast, our approach focuses on PIN inference without making any assumption on the similarity of keystroke acoustic signals. Another training-free approach~\cite{zhu2014context} takes full advantage of multiple pairs of microphones to estimate keystrokes' physical positions based on the time differences of keystroke acoustic signals arriving at different microphones. They require at least two pairs of microphones be placed near the users' keyboards so as to collect acoustic signals of users' keystrokes. In comparison, our attacks make no restriction on how to collect users' inter-keystroke timing data.
	
The second type of side-channel attacks for keystroke inference is video-based attack. Early works~\cite{backes2008compromising, backes2009tempest} exploit the reflections of screens on glasses, spoons, eyes of users to recover the users' inputs. These attacks require attackers to acquire videos directly capturing users' screens or screen reflections. Recent works show that even when keyboards or screens are not visible from the videos, attackers can still infer users' inputs via analyzing users' fingers or hands movements using advanced computer vision algorithms~\cite{yue2014blind, ye2017cracking, shukla2014beware}. Even users' hands movements are not visible from the videos, Sun et al.~\cite{sun2016visible} analyze the motion patterns of devices backsides caused by the users' keystrokes on different positions of the screens of the users' devices and classify them using Support Vector Machine. All these attacks require attackers to place cameras near the users.

The third type of side-channel attacks for keystroke inference is sensor-based attack. A range of studies show that embedded sensors on mobile devices or wearable devices can reveal sensitive information about users' keystroke behaviors. Various embedded sensors are investigated in this context, including accelerometers~\cite{marquardt2011sp, aviv2012practicality, liu2015good, maiti2016smartwatch, owusu2012accessory, wang2015mole}, gyroscopes~\cite{de2013identification, cai2011touchlogger, miluzzo2012tapprints, xu2012taplogger}, ambient-light sensors~\cite{spreitzer2014pin} and WiFi~\cite{ali2015keystroke, li2016csi}. All these attacks require attackers to hack into mobile devices or wearable devices for accessing sensor data or to place mobile devices near users' keyboards.
	
\section{Conclusion and Future Work}
In this paper, we proposed user-independent inter-keystroke timing attacks on PINs based on a human cognitive model. The human cognitive model allows an attacker to build a timing dictionary of all possible PINs ranked according to the cosine similarity between the observed timing sequence of a target PIN and each possible PIN's predicted timing sequence. We examined the effectiveness of our attacks to the PINs at different PIN strength levels in different online attack settings. The results demonstrated that our attacks achieve satisfactory performance. We also suggested several countermeasures to mitigate our attacks.

In the future, we plan to extend our attacks to infer users' passwords. A more complicated human cognitive model is to be developed for such purpose.

\section{Acknowledgements}
This research is supported by the Singapore National Research Foundation under NCR Award No. NRF2015NCR-NCR003-002. This research is also supported by AXA Research Fund. The work of Shujun Li was supported by the UK Engineering and Physical Sciences Research Council (EPSRC) under the grant number EP/N020111/1.

\bibliography{TimeDictionaryAttack}

\newpage
\appendix
\section{Other Cognitive Operations}
Apart from the four phenomena described in Section~\ref{phenomena}, we investigate other two cognitive operations which may affect the inter-keystroke time.

\textit{Phenomenon 5.} \textit{The execution process is slower for the transition from the last digit to the $<$Enter$>$ key.} People usually check the PIN before pressing the $<$Enter$>$ key, which makes the cognition time between the last digit and the $<$Enter$>$ key longer than the cognition time between the two keys in other key pairs.

\textit{Phenomenon 6.} \textit{Each character string can be decomposed either into syllables or trigrams and bigrams, sometimes even quatergrams.} People memorize words following this decomposition rule. In the processing of typing a string, a cognitive processor retrieves each segment of the string and encodes it into an ordered list of characters. The motor processor moves the finger and presses the key subsequently. The next segment of the string is then initiated and executed in the same way.
Specifically for PIN entry, people often break a long numeric code into smaller parts, and by `chunking' long numbers into several groups, people can greatly increase the recall accuracy and speed. For instance, it is common for people to memorize a 6-digit PIN following the `xxx-xxx' pattern which may lead to a slight pause after the first three numbers being entered~\cite{crook2008memory}. Sometimes, people use their birthdays to help define their PINs, so they memorize the PINs following the `yymmdd/yyddmm' or `mmddyy/ddmmyy' pattern. In these cases, the time between the segment `yy' and `mmdd/ddmm' maybe longer.

Phenomena 5 and 6 are two cognition operations of the typing task. In our paper, we call Phenomena 4 and 5 the word-end effect and the word-segment effect, respectively. We estimate the time of each operation as a constant parameter from experimental data.

Incorporated with the cognitive time (Phenomena 5 and 6), an extended linear model is used to predict the inter-keystroke time for 6-digit PINs:
\begin{equation}
\label{formula}
T = a + b * I + c * E + d * S_1 + e * S_2 + f * S_3,
\end{equation}
where the first two terms $a+b*I$ represent Fitts's Law, $E$ denotes the word-end effect, and $S_1, S_2, S_3$ denote the word-segment effect for three mostly-used PIN memorization patterns (`xx-xxxx', `xxx-xxx', `xxxx-xx' for 6-digit PINs), respectively. $E, S_1, S_2$ and $S_3$ are binary variables which take the value 1 if the input key pair is in the corresponding position in the PIN sequence (i.e., for the last key pair in a PIN sequence, $E=1$, otherwise, $E=0$; for the \textit{second}, \textit{third} and \textit{forth} key pair, $S_1, S_2, S_3 = (1, 0, 0)$, $(0, 1, 0)$, and $(0, 0, 1)$, respectively, otherwise, $S_1, S_2, S_3 = (0, 0, 0)$). The coefficients $c, d, e$, and $f$ denote the added time caused by such effects when they are present.

\begin{table}
	\centering
	\small
	\caption{The significance testing results of cognitive model's parameters. The statistically significant parameters are marked with $\bigstar$.}
	\begin{tabular}{c|l|c}
		\hline
		 Parameter  & Coefficient &   \textit{p}-value   \\ \hline\hline
		$Intercept$ & a = 155.08  & $<$ 0.001 $\bigstar$ \\ \hline
		    $I$     & b = 31.60   & $<$ 0.001 $\bigstar$ \\ \hline
		    $E$     & c = 17.63   &        0.060         \\ \hline
		  $S_{1}$   & d = 8.13    &        0.365         \\ \hline
		  $S_{2}$   & e = 7.81    &        0.429         \\ \hline
		  $S_{3}$   & f = 31.80   &        0.021         \\ \hline
	\end{tabular}
	\label{significance}
\end{table}

To validate this extended model, we conduct a significance testing on the coefficients of Equation \ref{formula}. The null hypothesis of each coefficient is $H_0: \beta = 0$ ($\beta$ refers to $c,d,e,f$ in Equation \ref{formula}, respectively). The results are given in Table \ref{significance}.

The results of significance testing show that the impact of word-end effect ($p=0.060$) and word-segment effects ($p=0.365;0.429;\\0.021$) are statistically insignificant while the topographical effect has a significant positive impact on the inter-keystroke time ($p<0.001$). One potential reason for the insignificant parameters (i.e., word-end and word-segment effects) is that typing a 6-digit PIN is a simple behavior so that participants may need little cognitive time for the PIN entry. Considering that the cognitive time is statistically insignificant, we decide to consider the topographical effect only in our cognitive model.

\end{document}